\documentclass[%
 amsmath,amssymb,
 aps,
notitlepage,
floatfix,
longbibliography
]{revtex4-1}

\usepackage{graphics}
\usepackage{dcolumn}
\usepackage{bm}

\usepackage[final]{changes}

\newcommand{\ub}{{\bf u}}

\newcommand{\Ub}{{\bf U}}
\newcommand{\Xb}{{\bf X}}

\newcommand{\xb}{{\bf x}}
\newcommand{\fb}{{\bf f}}
\newcommand{\Fb}{{\bf F}}

\newcommand{\dd}{{\mathrm d}}

\newcommand{\Rey}{\mathit{Re}}

\begin{document}

\preprint{APS/123-QED}

\title{Turbulence in a network of rigid fibers}

\author{Stefano Olivieri$^{1,2,3}$}
\email[Corresponding author:]{stefano.olivieri@oist.jp}
\author{Assad Akoush$^4$}
\author{Luca Brandt$^5$}
\author{Marco E. Rosti$^1$}
\author{Andrea Mazzino$^{2,3}$}
\affiliation{%
 $^1$ Complex Fluids and Flows Unit, Okinawa Institute of Science and Technology Graduate University, 1919-1 Tancha, Onna-son, Okinawa 904-0495, Japan\\
 $^2$ Department of Civil, Chemical and Environmental Engineering (DICCA), University of Genova, Via Montallegro 1, 16145, Genova (Italy) \\
 $^3$ INFN, Genova Section, Via Montallegro 1, 16145, Genova (Italy) \\
 $^4$ Department of Mechanical Engineering, Faculty of Engineering and Architecture, American University of Beirut, P.O. Box 11-0236, Riad El Solh, Beirut 1107 2020 (Lebanon) \\
 $^5$ Linn\'{e} FLOW Centre and SeRC, Department of Engineering Mechanics, KTH Royal Institute of Technology, Stockholm, Sweden
}

\date{\today}

\begin{abstract}
The effect of a network of fixed rigid fibers on fluid flow is investigated by means of three-dimensional direct numerical simulations using an immersed boundary method for the fluid-structure coupling. Different flows are considered (i.e., cellular, parallel and homogeneous isotropic turbulent flow) in order to identify the modification of the classic energy budget occurring within canopies or fibrous media, as well as particle-laden flows. First, we investigate the stabilizing effect of the network on the Arnold-Beltrami-Childress (ABC) cellular flow, showing that, the steady configuration obtained for a sufficiently large fiber concentration mimics the single-phase stable solution at a lower Reynolds number. Focusing on the large-scale dynamics, the effect of the drag exerted by the network on the flow can be effectively modelled by means of a Darcy's friction term. For the latter, we propose a phenomenological expression that is corroborated when extending our analysis to the Kolmogorov parallel flow and homogeneous isotropic turbulence. Furthermore, we examine the overall energy distribution across the various scales of motion, highlighting the presence of small-scale activity with a peak in the energy spectra occurring at the wavenumber corresponding to the network spacing. 
\end{abstract}

\maketitle



\section{\label{sec:intro}Introduction}

Scale separation is one of the most successful concepts exploited in a wide range of physical domains to derive large-scale dynamical descriptions where the small-scale dynamics only appear via effective parameters. The best known examples come from the solid state physics and kinetic theory. In the first field, the scale separation between the time scale associated to the motion of atomic nuclei and that of the electrons in a molecule allows one to treat them separately. As a result, the ion-ion interaction can be described by effective potentials on account of the motion of electrons. This is the essence of the well-known Born-Oppenheimer approximation.\\
In the gas kinetic theory, when the scales of interest are much larger than the mean free path (i.e., when the so-called Knudsen number is small), hydrodynamic equations can be derived where only macroscopic scales are involved, with the smallest involved only via effective parameters.\\
Scale separation  also constitutes the foundation for the emergence of effective parameters in the infra-red limit of fluid turbulence and turbulent transport: eddy-viscosity~\cite{dubrulle1991,germano1991,gama1994} and eddy-diffusivity~\cite{frisch1987lectures,frisch1995turbulence,biferale1995,mazzino1997,castiglione1998,mazzino2005} are well-known examples of effective parameters accounting for the non resolved small-scale dynamics.

Our aim here is to investigate how turbulence is modified when interacting with a  network of slender rigid fibers of length within the turbulence inertial range. Such a system can be seen as a rough model for quantitatively assessing how turbulence is modified by interacting with a canopy. This multiscale problem will be tackled by high-resolution DNS complemented by a state-of-the-art immersed boundary method (IBM) to fully resolve the fluid-structure interaction problem. Thanks to the full resolution of the whole coupled system, both the small-scale dynamics and the effective large-scale dynamics will be investigated. The latter will clearly emerge in the form of an effective Darcy equation when the flow field is observed on scales much larger than the typical correlation length of the solid network.

\begin{figure}
    \centering
    \includegraphics{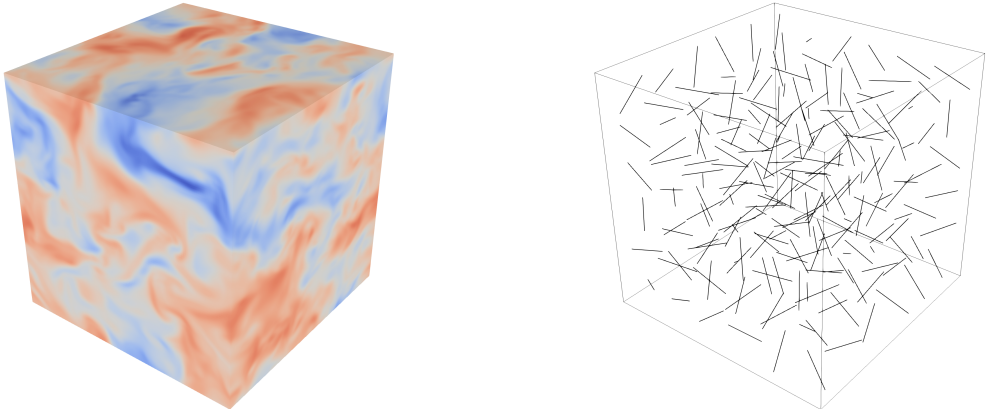}
    \caption{
    Left: homogeneous isotropic turbulent flow (the colormap showing one velocity component). Right: network of $N = 6^3$ fibers of length $c/L = (2 \pi)^{-1}$ placed in the same triperiodic fluid domain.}
    \label{fig:intro}
\end{figure}

One of the motivations of our study comes from boundary-layer meteorology where the interaction of wind with plant or urban canopies is known to cause modifications in the momentum and heat fluxes and velocity profiles~\cite{lemone100years}. Such features can influence the transport and mixing properties within the canopy, consequently altering ecological mechanisms such as carbon dioxide exchange~\cite{raupach1981}. Indeed, a relevant number of studies have been devoted to understand the underlying mechanisms by means of laboratory or in-situ measurements~\cite{poggi2004,DiBernardino2017,ghisalberti_nepf_2009a,shnapp2019extended}, as well as numerical simulations~\cite{zampogna_bottaro_2016,zampogna2016,monti2019PoF,monti2020JFM}. Detailed reviews on how turbulence is modified in plant canopies can be found in Refs.~\cite{raupach1981,finnigan2000review,nepf2012review}. \citet{finnigan2000review} highlighted, in particular, how the classical turbulent scenario, well explained by Kolmogorov theory~\cite{frisch1995turbulence}, can be substantially modified in the case of canopy flow. This is caused by the presence of canopy elements (e.g. twigs and leaves) exerting both viscous and pressure drag. According to~\citet{finnigan2000review}, the former has an overall dissipative effect, while the latter is responsible for generating wakes behind canopy elements, so that part of the large-scale kinetic energy is converted into smaller-scale kinetic energy (i.e., at higher wavenumbers). Such energy transfer was thus labeled as the `spectral shortcut' mechanism, which is believed to explain why the classical inertial subrange scaling $E(k) \sim k^{-5/3}$ does not hold within canopies~\cite{lemone100years}.

Such explanation, however, has been so far rather empirical and qualitative, lacking of a thorough analysis in a more quantitative framework. To gain a deeper comprehension, in the present work, the problem will be tackled in a more fundamental way, focusing on a relatively simple yet representative model: the fluid-solid interaction between a network of rigid fibers and several three-dimensional flows, i.e. (i) ABC cellular flow, (ii) Kolmogorov parallel flow, \added{(iii) homogeneous isotropic random flow}. Our three models aim to be an idealized but sufficiently general representation of both laminar flows with open and closed streamlines and a fully developed turbulent flow. In particular, the choice of the cellular and parallel flows will enable us to explore how the presence of fibers alters the stability properties compared to the single-phase case (i.e., without fibers), which has been already extensively investigated~\cite{galloway1987stability,podvigina1994}. The insights obtained in these two configurations will then be confirmed in the framework of homogeneous isotropic turbulence.
Additionally, we highlight that this kind of study has relevance also for the understanding of the interaction between fluid flows and dispersed fiber-like objects with large inertia, a topic of interest for many environmental/biological and industrial problems, such as aerosol deposition in the production of composite materials~\cite{duroure2019review}.

The rest of the paper is structured as follows: Sec.~\ref{sec:model} introduces the physical model and governing equations, along with describing how these are solved numerically; Sec.~\ref{sec:results} presents the results of our analysis and the main conclusions are drawn in Sec.~\ref{sec:conclusions}.

\section{\label{sec:model}Physical model and governing equations}

We consider a cubic domain with side $L$ and periodic boundary conditions, with an ensemble of approximately one-dimensional fibers immersed within, as shown in Fig.~\ref{fig:intro}. Each fiber composing the network has a length $c$ and diameter $d$ such that the aspect ratio $c/d \gg 1$. An integer number of fibers is placed along each direction with their centroids evenly spaced (one case of randomly spaced fibers will also be considered), while the orientation is always uniformly random distributed. The fiber concentration can be characterized by the number density $n = N/V$, where $N$ is the number of fibers and $V = L^3$ is the total volume of the fluid domain.

The fluid motion is governed by the incompressible Navier-Stokes equations
\begin{equation}
  \partial_{t} \ub + \ub\cdot\boldsymbol{\partial}\ub = -\boldsymbol{\partial} p/\rho_0 +\nu\partial^2\ub +  \fb,
  \label{eq:NS1}
\end{equation}
\begin{equation}
  \boldsymbol{\partial}\cdot\ub = 0,
  \label{eq:NS2}    
\end{equation}
where $\ub = (u,v,w)$ is the fluid velocity, $p$ the pressure, $\rho_0$ and $\nu$ the density and kinematic viscosity of the fluid and $\fb$ is the external volume forcing. The latter consists of two contributions, i.e. $\fb = \fb_\mathrm{FOR} + \fb_\mathrm{FIB}$: (i) $\fb_\mathrm{FOR}$ is a body force used to generate and sustain the desired flow, while (ii) $\fb_\mathrm{FIB}$ is the fluid-structure coupling term used to account for the presence of the fibers by an immersed boundary method (IBM). 

Specifically, we employ the same numerical procedure already used for moving and deforming filaments in laminar or turbulent flows in Refs.~\cite{rosti2018flexible,rosti2019flowing,banaei2019numerical,cavaiola2019assembly}, to which the reader is referred to for further information. The implementation relies on a finite difference, fractional step method on a staggered grid with fully explicit second-order central-differencing scheme in space and third-order Runge-Kutta scheme in time. Additionally, the Poisson equation enforcing the incompressibility constraint is solved using the Fast Fourier Transform.

The fluid-structure coupling force $\fb_\mathrm{FIB}$ is computed following the method by~\citet{huang_shin_sung_2007a} and later modified by~\citet{banaei2019numerical}. In particular, the Lagrangian force is first evaluated at each material point belonging to the fibers to enforce the no-slip condition $\Ub(\Xb(s,t),t)=\dot\Xb = \mathbf{0}$ as
\begin{equation}
    \Fb(s,t) = \kappa \, (\dot{\Xb} - \Ub),
\end{equation}
where $\kappa$ is a large negative constant \cite{huang_shin_sung_2007a} and 
\begin{equation}
 \Ub(\Xb(s,t),t) = \int \ub(\xb,t)\delta(\xb-\Xb(s,t))\,\dd\xb
\end{equation}
is the interpolated fluid velocity at the position $\Xb = \Xb(s,t)$ of the material point belonging to the fiber, as a function of the curvilinear coordinate $s$ and time $t$. A spreading is thus performed over the surrounding Eulerian points, yielding the volumetric forcing acting on the flow
\begin{equation}
   \fb_\mathrm{FIB}(\xb,t) = \int \Fb(s,t)\delta(\xb-\Xb(s,t))\, \dd s.
\end{equation}
Both the interpolation and spreading feature the Dirac operator, which is discretised by a regularized $\delta$; here, we employ the function proposed by~\citet{roma1999}.

In our simulations, we choose a domain size $L=2\pi$ and discretize into a Cartesian grid using $64$ points per side, with periodic boundary conditions applied in all directions. The fiber elements are discretized into $(N_\mathrm{L} - 1)$ segments with spatial resolution $\Delta s = c / (N_\mathrm{L} - 1)$, $N_\mathrm{L}$ being the number of Lagrangian points. Here we use $N_\mathrm{L}=11$ points, so that the Lagrangian spacing $\Delta s$ is approximately equal to the Eulerian grid size $\Delta x$. We verified that the variation of the results is negligible when doubling both the Eulerian and Lagrangian resolutions. As for the timestep, we use $\Delta t = 5 \times 10^{-5}$, after assessing convergence.

The described procedure has been implemented and extensively validated in both laminar and turbulent flow conditions: for more information the reader is referred to~\citet{rosti_brandt_2017a,rosti2018flexible,rosti2019flowing,banaei2019numerical}.

\section{\label{sec:results}Results}

\subsection{\label{sec:results_ABC}ABC cellular flow}

\begin{table*}
\caption{List of the cases for ABC cellular flows with a network of evenly spaced $N$ fibers of length $c$. $D$ here is the value measured from the decay of the energy spectrum.}
\label{tab:table1}
\begin{ruledtabular}
\begin{tabular}{lccccc}
$N$ & $c/L$ & $n$ & $D$ (measured) &　$\Rey_\mathrm{eff}$ &State \\
\colrule
0 & -- & 0 & -- & 130 & chaotic\\
$3^3$ & $(2\pi)^{-1}$ & 0.11 & -- & 53.9 & chaotic  \\
$4^3$ & $(2\pi)^{-1}$ & 0.26 & -- & 29.9 & chaotic \\
$5^3$ & $(2\pi)^{-1}$ & 0.50 & -- & 17.2 & pseudoperiodic \\
$6^3$ & $(2\pi)^{-1}$ & 0.87 & 0.11 & 10.5 & stable \\
$7^3$ & $(2\pi)^{-1}$ & 1.38 & 0.16 & 6.8 & stable \\
$8^3$ & $(2\pi)^{-1}$ & 2.06 & 0.19 & 4.6 & stable \\
$6^3$ & $(4\pi)^{-1}$ & 0.87 & -- & 16.0 & pseudoperiodic \\
$7^3$ & $(4\pi)^{-1}$ & 1.38 & 0.10 & 10.6 & stable \\
$8^3$ & $(4\pi)^{-1}$ & 2.06 & 0.14 & 7.3 & stable \\
$10^3$ & $(4\pi)^{-1}$ & 4.03 & 0.23 & 3.8 & stable \\
\end{tabular}
\end{ruledtabular}
\end{table*}

\begin{figure}
    \centering
    \includegraphics{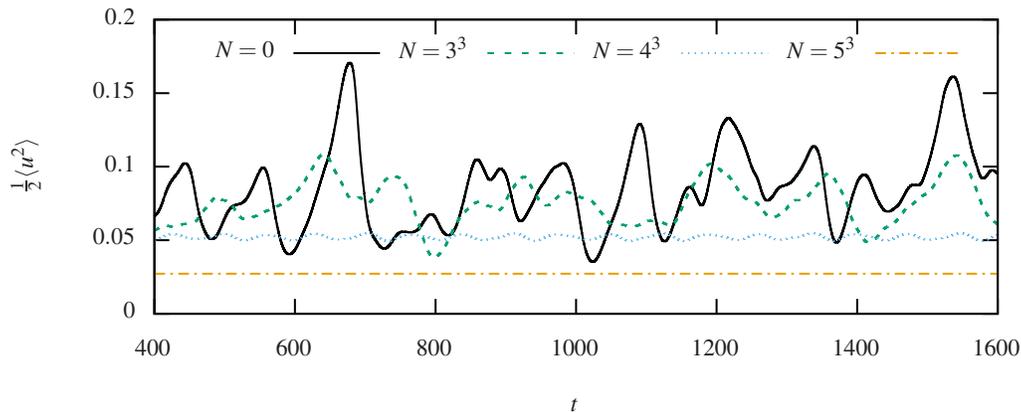}
    \caption{Time history of the mean fluid kinetic energy for different fiber concentrations in cellular ABC flow at $\Rey=130$.}
    \label{fig:k_vs_t_ABC}
\end{figure}

To start, we consider the so-called Arnold-Beltrami-Childress (ABC) flow~\citep{dombre1986chaotic}:
\begin{equation}
  \begin{aligned}
    u &= A \sin z + C \cos y\\
    v &= B \sin x + A \cos z\\
    w &= C \sin y + B \cos x
  \end{aligned}
  \label{eq:ABC}
\end{equation}
which is known to be a time-independent three-dimensional solution of both the Euler and Navier-Stokes equations, under the forcing  $\fb_\mathrm{FOR} = \nu( A \sin z + C \cos y, B \sin z + A \cos z, C \sin y + B \cos z)$, provided that the Reynolds number $\Rey \equiv \nu^{-1}$ is sufficiently small~\cite{arnold1965topologie,dombre1986chaotic,galloway1987stability}. The ABC flow is a special case of the Beltrami flows where the parameters $A$, $B$, and $C$ are real and the flow is periodic with respect to the Cartesian coordinates~\cite{zhao1993chaotic}. Despite its simple analytical expression, the flow provides an example of Lagrangian chaos~\citep{dombre1986chaotic,biferale1995eddy}. 
Stability analyses on this system have been carried out by several authors, see e.g. Refs.~\cite{galloway1987stability,podvigina1994}, which can therefore be used as a reference to study the effect of a network of fibers. In particular, choosing $A=B=C=1$, the flow becomes unsteady for $\mathit{Re} \geq 13$, showing an oscillatory behavior for $13 < \mathit{Re} \leq 20$ and becoming chaotic for $\mathit{Re} > 20$~\cite{podvigina1994}.

Here, we therefore fix $\Rey=130$ (for which in the absence of the dispersed phase the system is known to be unstable) and give to the initial condition a perturbation of finite amplitude at the same wavenumber of the forcing $k=1$. As a result, the solution given by Eq.~\eqref{eq:ABC} is lost in favour of an unsteady and chaotic regime~\cite{podvigina1994}. Using this setting, we conducted a parametric study by varying the number of fibers $N$ and the fiber length $c$, as presented in Table~\ref{tab:table1}.

Fig.~\ref{fig:k_vs_t_ABC} reports the time history of the mean fluid kinetic energy $\langle u^2 \rangle / 2$ (where $\langle \cdot \rangle$ denotes spatial average over the fluid domain) for cases with different concentration. Starting from the unstable flow without fibers (i.e. $N=0$), both the average value and the oscillation amplitude decrease for increasing $N$. Furthermore for sufficiently large concentrations ($N \geq 5^3$), we eventually reach a completely steady behavior. Thus, as expected, the presence of the fiber network stabilizes the flow provided that the network is sufficiently dense of fibers.

\begin{figure}
    \centering
    \includegraphics{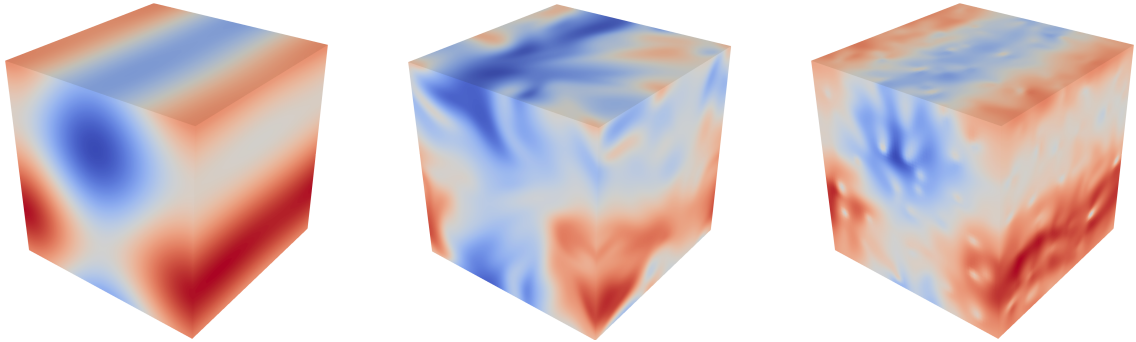}
    \caption{Visualizations of one component of the velocity field for three-dimensional cellular ABC flows. Left: stable solution given by Eq.~\eqref{eq:ABC}; center: unstable flow at $\Rey=130$ (at a given time instant); right: steady flow obtained in the presence of a network of $N=6^3$ evenly spaced fibers with $c/L = (2\pi)^{-1}$.}
    \label{fig:ABC_3d-vis}
\end{figure}

This result is qualitatively confirmed by the three-dimensional visualizations of the velocity field reported in Fig.~\ref{fig:ABC_3d-vis}. In the absence of the fiber network, the stable solution (left panel) given by Eq.~\eqref{eq:ABC} is obtained only if $\Rey < \Rey_\mathrm{cr}$, while we obtain the unsteady and chaotic solution for $\Rey > \Rey_\mathrm{cr}$ (center panel). Let us now focus on the case with fibers (right panel); as already pointed out, for a sufficiently dense network the resulting configuration is stable and steady. Looking at the velocity field, two main features can be observed: (i) the presence of wakes around the network elements, which can be associated to small-scale activity; (ii) the resemblance of the large-scale flow structure with that of the (stable) ABC flow.

\begin{figure*}
    \centering
    \includegraphics{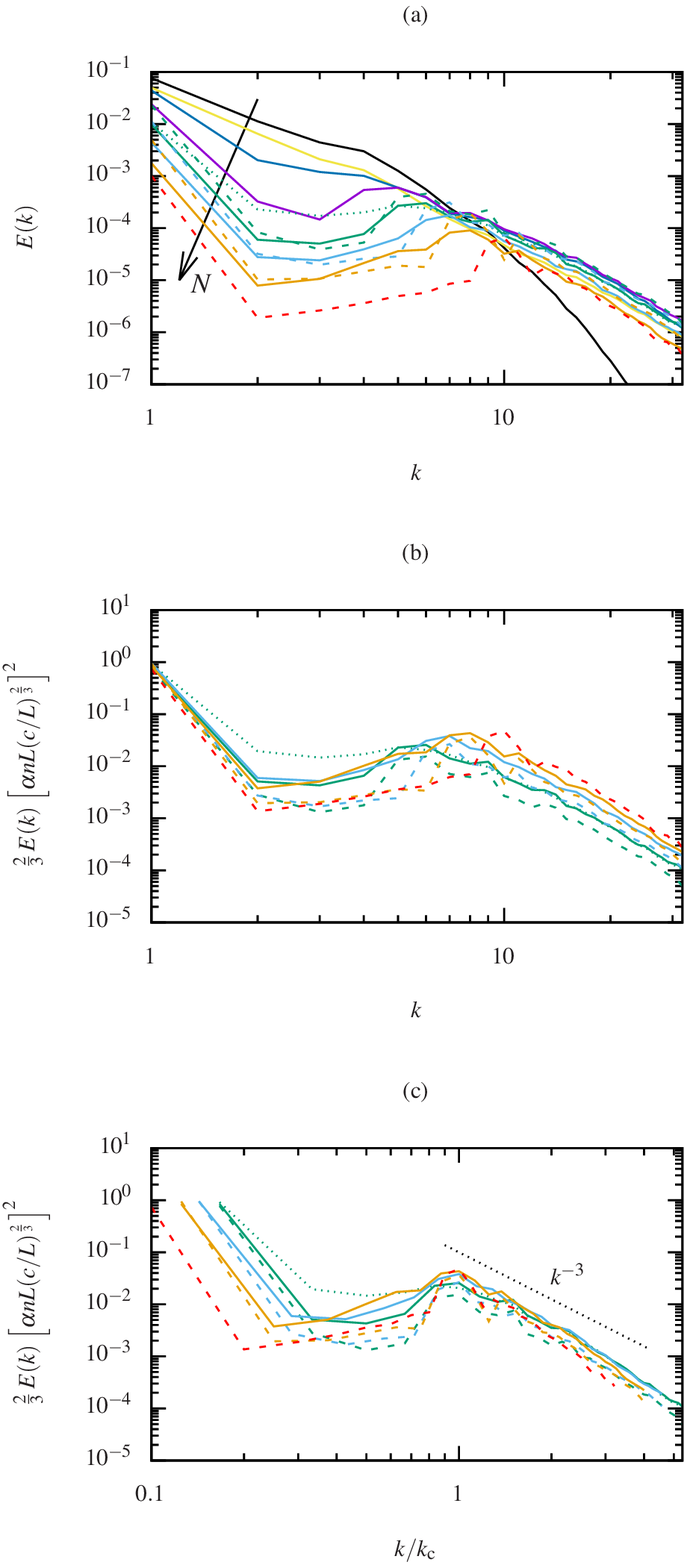}
    \caption{(a) Energy spectra in ABC cellular flow at $\Rey=130$ for different fiber concentrations (black: $N=0$, yellow: $N=3^3$, blue: $N=4^3$, violet: $N=5^3$, green: $N=6^3$, light blue: $N=7^3$, orange: $N=8^3$, red: $N=10^3$) and lengths (solid: $c/L=(2\pi)^{-1}$, dashed: $c/L=(4\pi)^{-1}$, dotted: $c/L=(2\pi)^{-1}$ with random positions).
    (b) Spectra of stabilized cases normalized using our argument based on Eqs.~\eqref{eq:u_rescaled} and \eqref{eq:exprD}.
    (c) Same but normalizing also the independent variable using the network wavenumber.}
    \label{fig:spectra_vs_nf}
\end{figure*}

Such evidence suggests to separate these two aspects by means of a scale-by-scale analysis. We have therefore computed the corresponding energy spectra, reported in Fig.~\ref{fig:spectra_vs_nf}a, from which it can be observed how the energy distribution across the scales of motion is modified by the presence of the network. We observe that, the energy associated to the large-scale/low-wavenumber components appears to decrease, while the small-scale/high-wavenumber activity becomes more relevant, consistently with what previously noted from the field visualization. One can notice that $k = 1$, the scale where the energy is introduced, remains always the dominant mode. For sufficiently high concentrations however, a secondary peak is seen to emerge. The wavenumber associated to this local maximum can be identified as $k_\mathrm{c} = 2 \pi / \ell = \sqrt[3]{N}$, $\ell$ being the characteristic lengthscale associated with the spacing between the network elements, i.e. $\ell = 2 \pi / \sqrt[3]{N}$. This scale is activated by virtue of the no-slip boundary condition imposed on each fiber. \added{
Looking at Fig.~\ref{fig:spectra_vs_nf}a, it is evident the shift of the secondary peak while varying the fiber concentration.
Moreover, it can be noted that the wavenumber related with the fiber length, i.e. $2 \pi / c$, is not associated with changes in the behavior of the energy spectrum, differently from what is typically claimed for turbulent flows within canopies~\cite{lemone100years,finnigan2000review}.}

\begin{figure}
    \centering
    \includegraphics{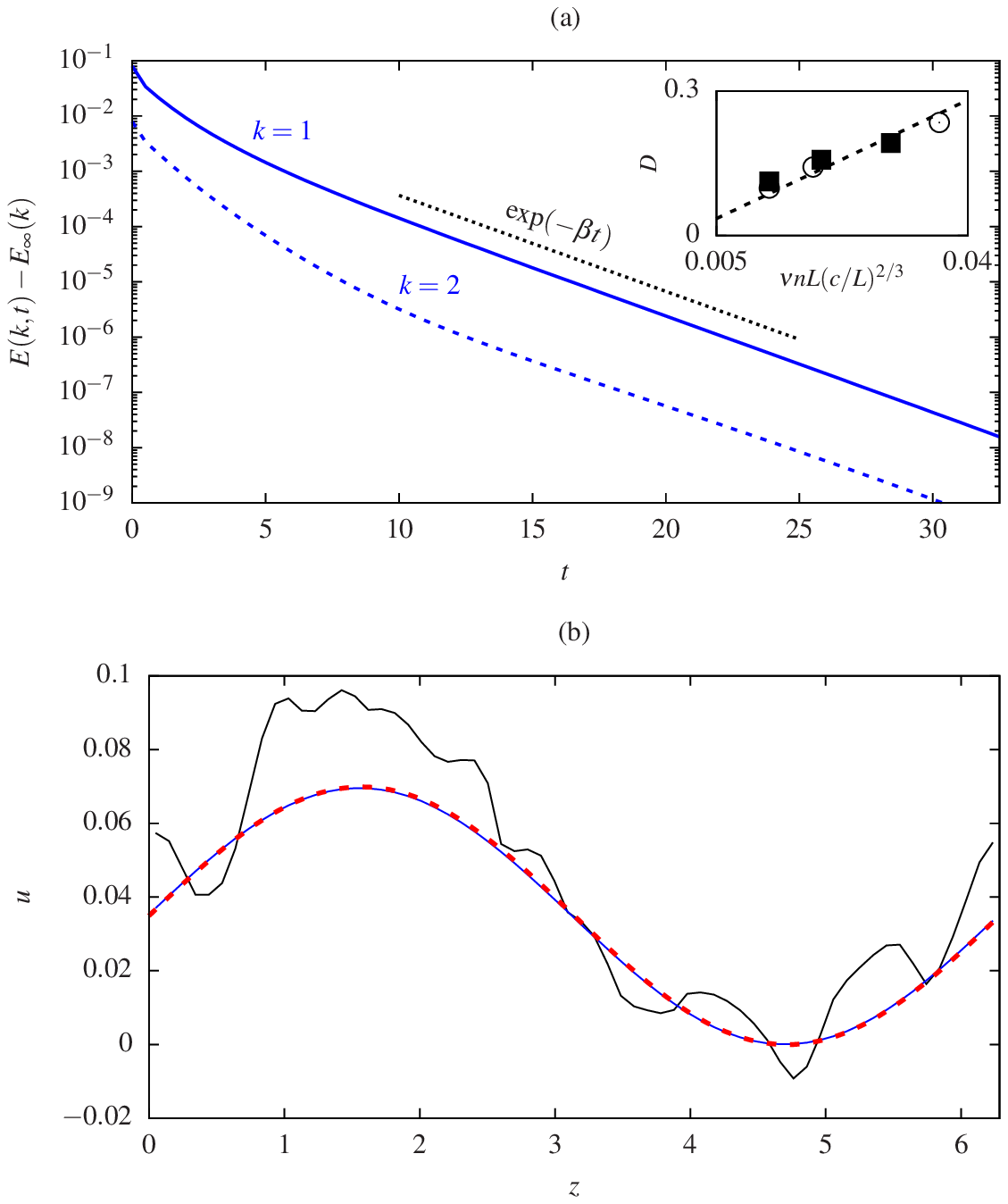}
    \caption{
    (a) Time history of the first and second mode of the energy spectrum for ABC flow at $\Rey=130$ and network of $N=8^3$ fibers with $c/L=(2\pi)^{-1}$ (solid blue line: $k=1$, dashed blue line: $k=2$, black dotted line: $\sim \exp(-\beta t)$); the inset reports the Darcy friction coefficient measured from our DNS (filled squares: $c/L=(2 \pi)^{-1}$, empty circles: $c/L=(4 \pi)^{-1}$) along with its expression proposed in Eq.~\eqref{eq:exprD} (dashed line).
    (b) Example of velocity profiles ($u$-component along $z$-direction) from the fully-resolved simulation (black solid line), the same but applying a large-scale filter (blue solid line) and that obtained using Eq.~\eqref{eq:u_rescaled}, i.e. the model equation for the large-scale motion.
    }
    \label{fig:spectra-decay_and_velocity-profile_ABC}
\end{figure}

The structure of the energy spectra discussed above, where a certain scale separation occurs, suggests the possibility of an effective description for the large-scale dynamics. To investigate this aspect, we look at how the low-wavenumber components of the energy spectrum decay in time. Here we focus on the case $N=8^3$ fibers, but similar findings are obtained for all the cases with sufficiently dense networks. The initial condition is a given frame from the fully-developed unstable configuration without fibers previously described. The time history of the first two components of $E(k=1,2)$ is shown in Fig.~\ref{fig:spectra-decay_and_velocity-profile_ABC}a. Except for the initial stages, the decay is substantially exponential for both modes, so that it can be expressed as $E(k,t) = E_0(k) \exp(-\beta t) + E_{\infty}(k)$, where $E_0$ and $E_{\infty}$ are the initial and asymptotic value (note that in the plot the spectrum is subtracted by the latter in order to highlight the exponential behavior), while $\beta$ is the characteristic time decay rate. From the energy balance, the governing equation for the energy spectrum can be written as follows:
\begin{equation}
    \partial_t {E(k,t)} =  T(k,t) + V(k,t) + F_\mathrm{FOR}(k,t) + F_\mathrm{FIB}(k,t),
    \label{eq:loads}
\end{equation}
with the various terms of the right-hand-side corresponding to the nonlinear energy transfer, the viscous dissipation, the external flow forcing and the fluid-structure coupling, respectively; for the definition of these quantities, see Appendix~\ref{app:spectral-balance}. For the moment, let us neglect the nonlinear term (this assumption will be later justified).

From Fig.~\ref{fig:spectra-decay_and_velocity-profile_ABC}a, we can observe that the first and second spectral mode decay essentially at the same rate (in this case $\beta \approx 0.4$). This indicates that, when focusing on the large-scale dynamics, the effect of the network of fibers can be modelled by means of a Darcy-like friction term $F_\mathrm{FIB}(k,t) = - D \, E(k,t)$, where $D$ represents the friction factor. Indeed, the decay associated with such friction term turns out to be independent from the wavenumber $k$. Conversely, for the viscous dissipation $V(k,t) = - 2 \nu k^2 E(k,t)$ it scales as $k^2$. Considering both contributions, the overall decay rate is thus $\beta = 2 (D + \nu k^2)$. Testing this assumption against our DNS data, the effective Darcy friction is found to be dominant compared with the viscous dissipation, resulting in the observed independence of $\beta$ from the spectral mode $k$.

We thus focus only on the largest scale, i.e. on the first wavenumber $k=1$, and shift our attention back to physical space, where the fluid-structure forcing can be modelled now as $\fb_\mathrm{FIB} = -D \ub$. Consequently, we can write a balance between the resulting large-scale velocity field $\ub^\mathrm{eff}$ (governed by the external forcing, the viscous and the Darcy terms) and the single-phase solution of the ABC flow $\ub^\mathrm{ABC}$ (Eq.~\eqref{eq:ABC}) (governed only by the forcing and the viscous terms) as 
\begin{equation}
    D \ub^\mathrm{eff} - \nu \partial^2 \ub^\mathrm{eff} + \fb_\mathrm{FOR} = -\nu \partial^2 \ub^\mathrm{ABC} + \fb_\mathrm{FOR}.
    \label{eq:ABC_eff}
\end{equation}
Combining Eqs.~\eqref{eq:ABC_eff} and~\eqref{eq:ABC}, the large-scale velocity field $\ub^\mathrm{eff}$ is thus simply obtained as
\begin{equation}
    \ub^\mathrm{eff} = \frac{\nu}{\nu + D} \, \ub^\mathrm{ABC},
    \label{eq:u_rescaled}
\end{equation}
i.e., the large-scale flow is the same as the ABC solution, although with a reduced amplitude compared to the one obtained without fibers. Note that, this is formally valid only when the nonlinear terms are small so that they can be neglected in the balance.

To validate our argument, Fig.~\ref{fig:spectra-decay_and_velocity-profile_ABC}b reports a sample of the resulting velocity profile, which can be filtered to remove the small-scale components retaining only the first modes. Comparing the latter with the profile given by Eq.~\eqref{eq:u_rescaled}, where we highlight that $D$ is measured from the decay of the energy spectrum, we find that the agreement is very good, confirming the validity of our approach. Such procedure has been applied to the other five cases reported in Table~\ref{tab:table1} for which a stable state is reported, with the aim of deducing an expression of $D$ as a function of the main parameters involved in the problem. \added{From the numerical evidence, we find on empirical basis that the Darcy's coefficient can be written as
\begin{equation}
    D = \alpha \, \nu \, nL \, \left( \frac{c}{L} \right)^{\frac{2}{3}},
    \label{eq:exprD}
\end{equation}
where $\alpha \approx 7$ is a dimensionless factor found by fitting this expression to our data. Note that, overall this expression resembles the typical structure of the Darcy's term used for porous media models~\cite{bottaro2019review}, although we have the dimensionless correction $(c/L)^{2/3}$ which is purely phenomenological. Even if we cannot offer a theoretical support for such correction, the accuracy of Eq.~\eqref{eq:exprD} can be detected from the inset of Fig.~\ref{fig:spectra-decay_and_velocity-profile_ABC}a, where the results obtained for networks with different $N$ and/or $c/L$ collapse reasonably well. The form of Eq.~\eqref{eq:exprD} deserves some comments: i) both the fluid viscosity $\nu$ and the network concentration $n$ enter linearly into the expression, similarly to what is found for the classical Darcy's term~\cite{bottaro2019review}; ii) the box size $L$ is used as the characteristic length representative of the elementary cell volume; iii) the dependence of $D$ on the fiber length $c$ is found to be weaker than linear.} 

Using again the similarity with the stable ABC solution (Eq.~\eqref{eq:u_rescaled}), we can evaluate an effective Reynolds number,  $\Rey_\mathrm{eff} = (\nu + D)^{-1}$, on the basis of which the fluid flow can be characterized. To be consistent with the results of the stability analysis for the classical ABC flow, $\Rey_\mathrm{eff} \lesssim \Rey_\mathrm{cr}$, where $\Rey_\mathrm{cr} \approx 13$ is the aforementioned critical value for the first instability. In Table~\ref{tab:table1}, we report the value of $\Rey_\mathrm{eff}$ along with the observed state (e.g., stable or chaotic) for each case: the correlation between the two is as expected, further confirming that the approach here proposed can effectively model the presence of the network.

Moreover, the magnitude of the first mode of the energy spectrum can be derived from Eq.~\eqref{eq:u_rescaled}. Indeed, in the classical ABC flow the latter is equal to $3/2 A^2$, while in the presence of fibers we have $3/2 A'^2$, with $A' = \nu / (\nu + D) \approx [\alpha n L (c/L)^{2/3}]^{-1}$ (neglecting the viscous term contribution). Using this quantity, we normalize the energy spectra (of those cases where $\Rey_\mathrm{eff} < \Rey_\mathrm{cr}$) as shown in Fig.~\ref{fig:spectra_vs_nf}b, and show that the large-scale/low-wavenumber components substantially overlap. Moreover, the energy spectra can be normalized also in the independent variable using the network wavenumber $k_\mathrm{c}$ discussed previously, as depicted in Fig.~\ref{fig:spectra_vs_nf}c where one can notice that all curves are collapsing for $k/k_\mathrm{c} \geq 1$, i.e. in the small-scale range. In this range the scaling resembles a power law $\sim k^{-3}$, the fingerprint of a regime having smooth fluctuations in space.

\begin{figure}
    \centering
    \includegraphics{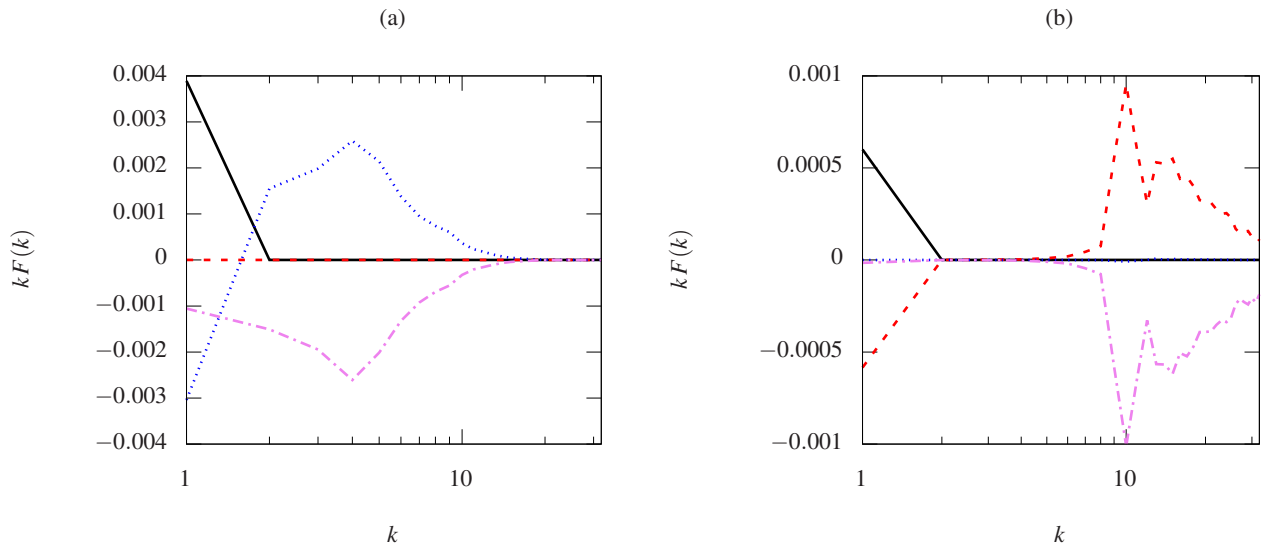}
    \caption{Spectral power balance according to Eq.~\eqref{eq:loads}, multiplied by $k$ to improve the plot readibility, for cellular ABC flow at $\Rey=130$: (a) unstable case without fibers ($N = 0$), and (b) stabilized flow for a network of $N=10^3$ fibers with $c/L= (4 \pi)^{-1}$. Black solid line: external forcing; red dashed: fluid-structure coupling; blue dotted: nonlinear term; violet dot-dashed: viscous dissipation. The vertical dashed line in (b) indicates the network wavenumber $k_\mathrm{c}$.}
    \label{fig:loads_ABC}
\end{figure}

One important aspect has to be underlined regarding the limit of validity of Eq.~\eqref{eq:u_rescaled}: as we already stated, the balance that we considered, between the effective friction and the external forcing, relies on the fact that the fiber network is sufficiently dense so that the friction is large enough and the nonlinear terms in the Navier-Stokes equations can be neglected (the latter will be discussed later together with Fig.~\ref{fig:loads_ABC}b).

To complete the analysis, we consider again the spectral energy budget in Eq.~\eqref{eq:loads} and compute from our numerical data each term appearing on the right-hand-side, reported in Fig.~\ref{fig:loads_ABC} for the unstable flow and one case stabilized by the fiber network. We first consider the single-phase case (Fig.~\ref{fig:loads_ABC}a). As prescribed, the external forcing acts only on the first mode. The energy input is balanced only partially by dissipation, while the remaining is transferred by the nonlinear term to higher wavenumbers. Note that here the flow is chaotic but not properly turbulent, due to the limited $\Rey$, so that no characteristic energy cascade can be observed. Nevertheless, although limited to few modes, a certain proliferation of active scales of motion occurs (see again the black curve in Fig.~\ref{fig:spectra_vs_nf}a), up to the condition where viscous dissipation becomes dominant.\\
Let us now move to the case with a network of $N=10^3$ fibers with $c/L=(4 \pi)^{-1}$ (Fig.~\ref{fig:loads_ABC}b). While the external forcing has obviously the same structure as before, the energy input from the external forcing is reducing (compared to the case without fibers) because of the reduced mean kinetic energy (as it was shown in Fig.~\ref{fig:k_vs_t_ABC}). Remarkably, the nonlinear term is found to be negligible for every $k$ if compared to the other terms, thus justifying the assumption previously made. Focusing on $k=1$, it is evident how the large-scale dynamics is given by a balance between the forcing and the fluid-structure coupling: the former injects energy in the system, while the latter subtracts it. Furthermore, Fig.~\ref{fig:loads_ABC}b also shows a positive peak for the fluid-structure coupling at $k = k_\mathrm{c}$ that is substantially balanced by viscous dissipation. The same behavior occurs at higher wavenumbers $k > k_\mathrm{c}$, although with the tendency for both terms to decrease with $k$. We also point out that all terms are vanishing over an intermediate range of wavenumbers, approximately $2 \leq k \leq 6$, indicating once again the nonlocal energy transfer operated by the fiber network.

To summarize, the analysis carried out on the cellular flow reveals the stabilizing effect of a sufficiently dense network of fibers, which can be effectively described using a Darcy-like friction term when focusing on the larger scales of motion. At smaller scales, a nonlocal energy transfer mechanism is responsible of a secondary peak in the energy spectrum emerging at a lengthscale associated with the spacing between the network elements. Before concluding this section, it is worth noticing that the whole analysis holds also when the fibers are arranged completely random. In this case, as shown in Fig.~\ref{fig:spectra_vs_nf} with dotted lines, the spectra still present the same features previously described and the only difference is a broader secondary peak than in the ordered case, being the lengthscale associated with the spacing between the network elements not uniquely defined anymore. In the following sections, we aim to assess whether the proposed phenomenological model still works in different flows.

\subsection{\label{sec:results_Kol}Kolmogorov flow}

As a complementary case to the cellular flow previously examined, we now consider a parallel flow configuration. In particular, we choose the so-called Kolmogorov flow, defined as~\cite{meshalkin1961}
\begin{equation}
  \begin{aligned}
    u &= \cos y\\
    v &= 0\\
    w &= 0,
  \end{aligned}
  \label{eq:Kol}
\end{equation}
where the streamlines are open and the only nonzero velocity component varies sinusoidally along the transverse direction. To obtain the solution above (provided that $\Rey < \Rey_\mathrm{cr}$), the external forcing in Eq.~\eqref{eq:NS1} is now expressed as $\fb_{\mathrm{FOR}} =  - \nu \, \cos y \, \mathbf{e}_x$. Similarly to the ABC flow, this configuration represents a prototype for the investigation of hydrodynamic stability, both for Newtonian and non-Newtonian fluids~\cite{thess1992,boffetta2005viscoelastic,boffetta2005drag,bistagnino2007nonlinear,tithof2017bifurcations}. In the single-phase case, the critical value of the Reynolds number is found theoretically to be $\Rey_\mathrm{cr} = \sqrt{2}$~\cite{meshalkin1961,thess1992} and the transition from the stable solution given by Eq.~\eqref{eq:Kol} occurs only if the flow is perturbed on a lengthscale much larger than that of the base flow~\cite{boffetta2005viscoelastic}.
In the absence of scale separation, i.e. the ratio between the perturbation and the base flow wavenumber is $\mathcal{O}(1)$, $\Rey_\mathrm{cr}$ is slightly larger than $\sqrt{2}$. 
Accordingly, we fix $\Rey=100$ and double the domain size to $2 L = 4 \pi$, imposing a low-wavenumber initial perturbation on $k=1/2$. In this setting, the flow turns out to be unstable. We let the flow evolve up to $t=500$ and then add the fiber network. In particular, we have performed simulations considering two different concentrations, $N=4^3$ and $7^3$, with fiber length equal to $c/L=(2 \pi)^{-1}$. In addition, to test the validity of our simple model, we have conducted the same simulations replacing into the Navier Stokes equations the fully-resolved IB approach for the fiber network with the effective Darcy's term, i.e. $\fb_\mathrm{FIB} = - D \ub$, where $D$ is given by Eq.~\eqref{eq:exprD} (without adjusting the free parameter $\alpha$ in the expression).

\begin{figure}
    \centering
    \includegraphics{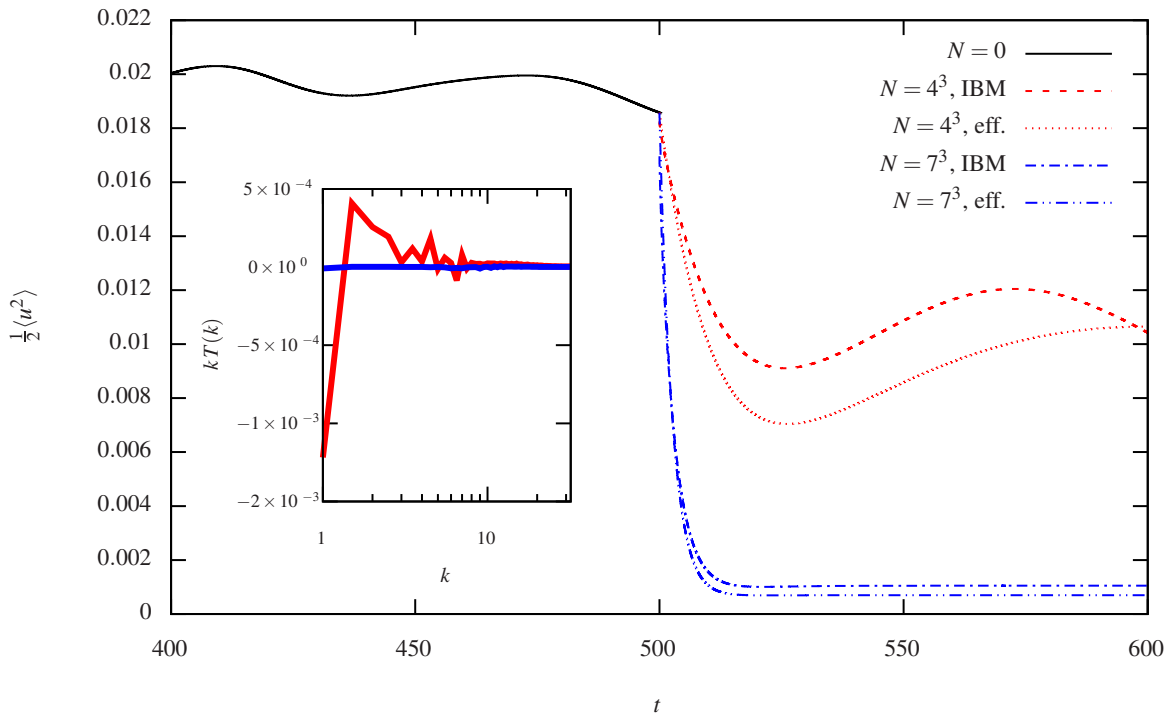}
    \caption{Time history of mean fluid kinetic energy for different fiber concentrations in a Kolmogorov flow at $\Rey=100$. Both the results from the fully-resolved simulations and the large-scale effective model are reported. Inset: nonlinear terms appearing into the spectral balance Eq.~\eqref{eq:loads} for the cases with fibers.}
    \label{fig:k_vs_t_Kol}
\end{figure}

The time history of the kinetic energy is reported in Fig.~\ref{fig:k_vs_t_Kol} for different fiber concentrations. The same phenomenology already observed for the ABC flow can be recognized, with the overall stabilizing role of the network and, in particular, a steady solution obtained for the highest concentration that has been tested. It is also important to note that the behavior of the system forced only at the large scale by a friction Darcy-like term is in good agreement with the results from the fully resolved simulations. The effective description is thus capable of capturing correctly the dynamical state reached by the flow under the action of the fluid-structure coupling. 
Moreover, one can see that the agreement improves for increasing concentrations; again, this is consistent with the fact that the nonlinear terms, whose contribution to the spectral energy budget is shown in the inset of Fig.~\ref{fig:k_vs_t_Kol}, decrease while increasing $N$, thus making the validity of our model stronger.

\begin{figure}
    \centering
    \includegraphics{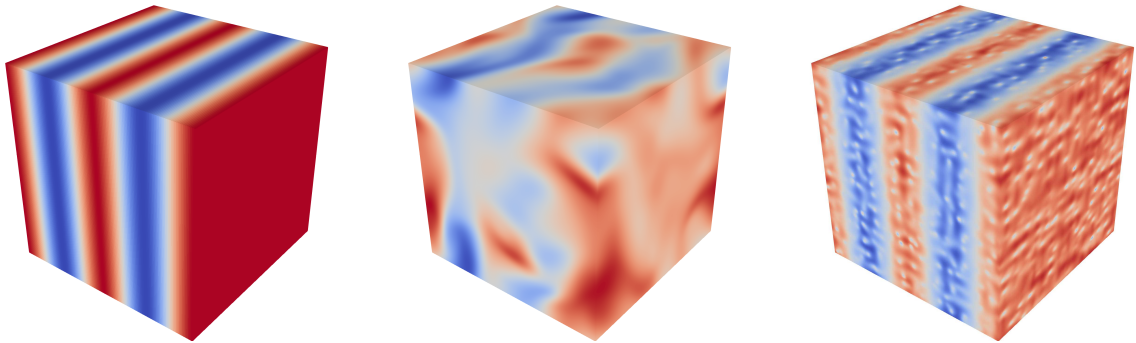}
    \caption{Visualizations of one component of the velocity field for the Kolmogorov parallel flows. Note that the fluid domain size is now $2 L = 4 \pi$. Left: stable solution given by Eq.~\eqref{eq:Kol}; center: unstable flow at $\Rey=100$ (at a given time instant); right: steady flow obtained in the presence of a network of $N=7^3$ evenly spaced fibers with $c/L = (2\pi)^{-1}$.}
    \label{fig:Kol_3d-vis}
\end{figure}

Focusing on the steady case with higher fiber concentration, in Fig.~\ref{fig:Kol_3d-vis} we compare the resulting velocity field, similarly to what done previously for the cellular flows. While in the absence of fibers, for $\Rey > \Rey_\mathrm{cr}$, the stable Kolmogorov flow solution (left panel) is lost to reach the unstable one (center panel), in the presence of a sufficiently dense fiber network the fingerprint of the stable solution is recovered in the resulting large-scale configuration (right panel).
Finally, we compare the single-phase solution obtained at a Reynolds number equal to the effective Reynolds number of the fiber-laden case (as previously done in Sec.~\ref{sec:results_ABC}), verifying that a similar steady solution is recovered. 

\subsection{\label{sec:results_HIT}Stochastic forcing}

As the final step of our analysis, we consider a widely used model for turbulent flows: a homogeneous isotropic flow subject to a stochastic forcing (at high Reynolds number, this is usually called homogeneous isotropic turbulence)~\cite{frisch1995turbulence}. To reproduce numerically this configuration, the flow is sustained using the spectral forcing scheme by~\citet{eswaran1988forcing}, where energy is injected randomly at low wavenumbers (in our case, within a spherical shell with radius $k=2$) by means of a Ornstein-Uhlenbeck process.

We choose a setting where, in the absence of fibers, the Reynolds number based on the Taylor's microscale is $\Rey_\lambda \approx 40$, and consider two different fiber concentrations, $N=4^3$ and $10^3$, with $c/L=(4 \pi)^{-1}$. As before, we have simulated the flow without fibers up to a certain time $t=100$, then used as the initial condition for the cases with fiber network. Furthermore, we also carried out computations using the effective model for the same parameter setting, based again on Eq.~\eqref{eq:exprD} for the Darcy's coefficient.

\begin{figure}
    \centering
    \includegraphics{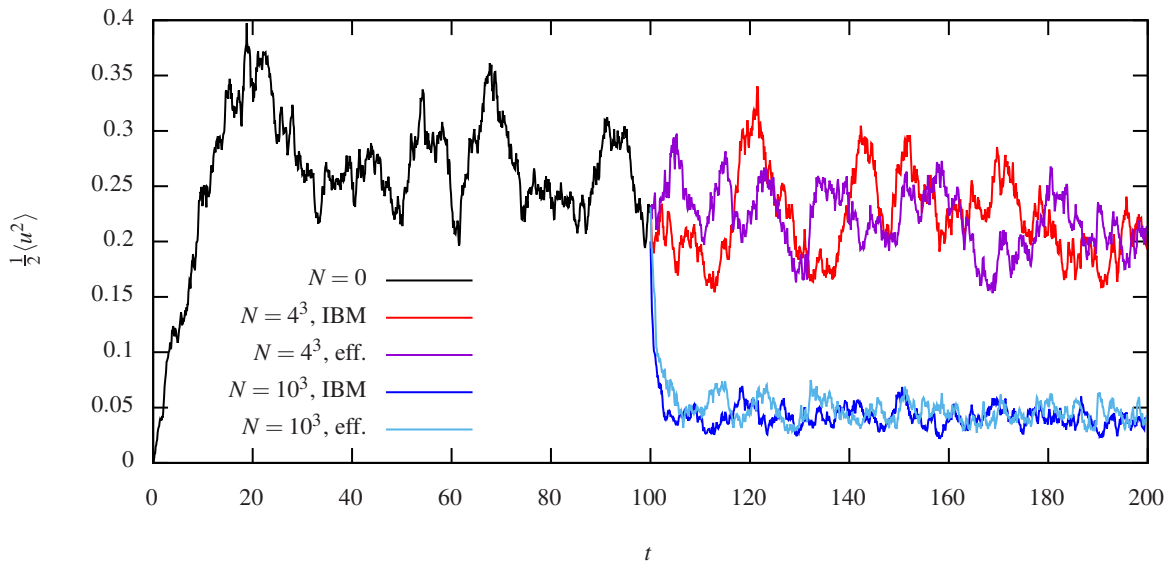}
    \caption{Time history of the mean fluid kinetic energy for different fiber concentrations in homogeneous isotropic turbulence at $\Rey_\lambda \approx 40$. Both the results from the fully-resolved simulations and the large-scale effective model are reported.}
    \label{fig:k_vs_t_HIT}
\end{figure}

\begin{figure}
    \centering
    \includegraphics{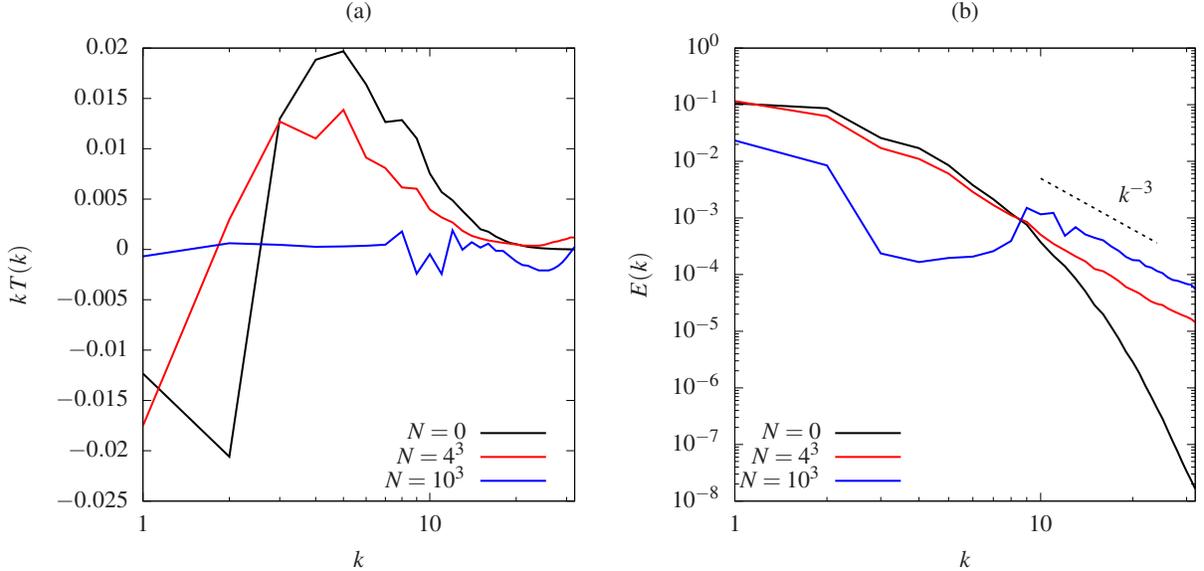}
    \caption{(a) Nonlinear term contribution of the spectral power balance according to Eq.~\eqref{eq:loads} and (b) energy spectra in homogeneous isotropic turbulence at $\Rey_\lambda \approx 40$ for different fiber concentrations (black: $N=0$, red: $N=4^3$, blue: $N=10^3$) with fiber length $c/L=(4\pi)^{-1}$. Both the results from the fully-resolved simulations and the large-scale effective model are reported. The vertical dotted lines indicate the corresponding network wavenumbers.}
    \label{fig:loads_and_spectra_HIT}
\end{figure}

First, we look at the kinetic energy in time, reported in Fig.~\ref{fig:k_vs_t_HIT}. For the less concentrated case with $N=4^3$ fibers, the effect of the network appears to be very limited, while for $N=10^3$ we have a significant reduction of the average value, along with a decreased oscillation amplitude. Nevertheless, despite the stabilizing effect of the fiber network, the flow always remains unsteady, clearly because of the random forcing that is applied. Comparing on a statistical basis the large-scale observables from the fully-resolved cases with those from the corresponding effective ones, good agreement is found for both concentrations (note that the comparison is limited to the large-scale dynamics since the effective simulations are not able to describe the full range of wavenumbers and therefore reproduce, e.g., the energy spectrum).
 To check the consistency with the picture already drawn for the ABC and Kolmogorov flows, we look once again at the role of the nonlinear terms in the spectral balance, Eq.~\eqref{eq:loads}, shown in Fig.~\ref{fig:loads_and_spectra_HIT}a: the same trend observed before for the Kolmogorov flow (see the inset of Fig.~\ref{fig:k_vs_t_Kol}) is found also here, with $T(k)$ decreasing when increasing $N$.

In this regard, we point out that this scenario can change for a different choice of the governing parameters. For example, we have conducted some tests (not shown here) where the viscosity is decreased and, in turns, $T(k)$ becomes more important with respect to the other terms in Eq.~\eqref{eq:loads}.
In such conditions, the applicability of a Darcy-like description is not sufficient anymore and the need of a more elaborated model taking into account the role of inertial terms is required.

Next, the energy spectra are shown in Fig.~\ref{fig:loads_and_spectra_HIT}b. Overall, one can note the close resemblance with the trend observed when considering the ABC cellular flow (Fig.~\ref{fig:spectra_vs_nf}a). On one hand, the large-scale/low-wavenumber components decrease for increasing the fiber concentration, while the opposite occurs for the small-scales/high-wavenumbers, for which we recover the same scaling $\sim k^{-3}$, fingerprint of smooth-in-space velocity excursions. Furthermore, for the case with the highest concentration we can clearly observe a region of low energy for intermediate wavenumbers and the peak at $k_\mathrm{c} = 10$, representing the signature of the non-local mechanism of energy distribution previously identified.
Finally, Fig.~\ref{fig:loads_and_spectra_HIT}b shows also the energy spectra obtained from the effective-model simulations, which as expected compare well with the fully-resolved simulations at the large scales while do not capture the small-scale information.

\section{\label{sec:conclusions}Conclusions}

We perform direct numerical simulation of three periodic flows with a network of fixed rigid fibers suspended within; the presence of the fiber is simulated using an immersed boundary method which handles the fluid-structure coupling. In particular, the Arnold-Beltrami-Childress cellular flow with closed streamlines, the parallel Kolmogorov flow with open streamlines and the homogeneous isotropic turbulent flow are considered in order to understand the stability and modifications of energy transfer of flows within canopies, fibrous media, and particle-laden flows.

First, we find that the fiber network has a stabilizing effect on the flows. Indeed, the ABC flow can be stationary even at large Reynolds numbers in the presence of a high concentration of fibers, with the resulting stationary flow mimicking the single phase stable solution at a lower Reynolds number. Based on this evidence, we therefore perform separate analysis for the large and small scales of the flow. For the large-scale dynamics, we find that the effect of increased drag exerted by the network of fibers on the flow can be effectively modelled by means of a Darcy's friction term. This can be used  to model the large scale motion of the flow and is tested in all the flows considered here, i.e. the ABC, Kolmogorov and turbulent flows. As concerns the small-scale dynamics, we find that the presence of fibers triggers small-scale activity, which results in an energy spectrum with the emergence of a secondary peak at a wavenumber corresponding to the
the network spacing. By examining the overall energy distribution across the various scales of motion, we find that the non-linear contribution to the energy balance rapidly vanishes as the concentration of the network grows, with the fluid-structure coupling term balancing the external forcing at the large-scales and the viscous dissipation at the small-scales. The fluid-structure coupling dissipates energy at the large scale and re-introduces energy in the system at the small ones, thus effectively acting as a nonlocal energy transfer mechanism.

This work highlights the key features of a fiber network on the flow. Our analysis clarified the origin of the modifications of the energy spectrum in the presence of suspended rigid fibers, which however are common in several other systems, ranging from canopy flows, flows in porous media and even suspension flows. At which extent our findings apply when the fibers are freely moving into the flow is the subject of future investigations.

\acknowledgments
SO acknowledges OIST for supporting his visiting period in the Complex Fluids and Flows Unit. AM thanks the financial support from the Compagnia di San Paolo, project MINIERA n. I34I20000380007. LB acknowledges financial support from the Swedish Research Council (VR), Grant No.\ VR 2014-5001. Computing time was provided by INFN and CINECA.

\appendix

\section{Energy spectrum equation}
\label{app:spectral-balance}

In this appendix, we briefly recall how Eq.~\eqref{eq:loads} is derived, along with identifying each term appearing in the equation. A detailed explanation can be found in classical textbooks, see e.g.~\citet{pope2000turbulent}.

As the starting point, we perform the Fourier transform of the Navier-Stokes Eqs.~\eqref{eq:NS1} and~\eqref{eq:NS2}, yielding:
\begin{equation}
  \partial_t \hat{\ub} + \hat{\mathbf{G}} = - i \mathbf{k} \hat{p}/\rho_0 - \nu k^2 \hat{\ub} + \hat{\fb},
  \label{eq:NS1_spectr}
\end{equation}
\begin{equation}
  \mathbf{k} \cdot \hat{\ub} = 0,
  \label{eq:NS2_spectr}    
\end{equation}
where $\hat{(\cdot)}(\mathbf{k},t) = \mathcal{F}\{ (\cdot)(\xb,t) \}$ denotes the Fourier transform, $\mathbf{G}$ corresponds to the nonlinear term appearing in the momentum equation and $\mathbf{k}$ is the wavenumber vector. The same equations can be written for the complex conjugate $\hat{\ub}^*$. Multiplying Eq.~\eqref{eq:NS1_spectr} by $\hat{\ub}^*$, the pressure term drops due to the incompressibility constraint, Eq.~\eqref{eq:NS2_spectr}, and the same applies in the momentum equation for $\hat{\ub}^*$ when multiplying by $\hat{\ub}$.

When summing the two equations for $\hat{\ub}$ and $\hat{\ub}^*$, we obtain an equation for the spectral kinetic energy $\hat{E}(\mathbf{k},t)$, defined as $\hat{E}(\mathbf{k},t) = \langle \hat{\ub}^* \cdot \hat{\ub} \rangle / 2$, which reads as
\begin{equation}
  \partial_t \hat{E} = \hat{T} + \hat{V} + \hat{F},
  \label{eq:spectral_energy}
\end{equation}
where we have identified the following quantities:
\begin{itemize}
    \item[--] $\hat{T} = \frac{1}{2} \, (\hat{\mathbf{G}} \cdot \hat{\ub}^* + \hat{\mathbf{G}}^* \cdot \hat{\ub}) $ is the transfer term associated with the nonlinear convective term;
    \item[--] $ \hat{V} = - 2 \nu k^2 \hat{E} $ is the viscous dissipation;
    \item[--] $\hat{F} = \frac{1}{2} \, (\hat{\fb} \cdot \hat{\ub}^* + \hat{\fb}^* \cdot \hat{\ub})$ is the energy input/output associated with the forcing.
\end{itemize}
Finally, to obtain Eq.~\eqref{eq:loads} for the energy spectrum $E(k,t)$, Eq.~\eqref{eq:spectral_energy} is averaged in each direction for isotropy, i.e. over a sphere of radius $k$.


\bibliography{ms}

\providecommand{\noopsort}[1]{}\providecommand{\singleletter}[1]{#1}%
\begin{thebibliography}{43}%
\makeatletter
\providecommand \@ifxundefined [1]{%
 \@ifx{#1\undefined}
}%
\providecommand \@ifnum [1]{%
 \ifnum #1\expandafter \@firstoftwo
 \else \expandafter \@secondoftwo
 \fi
}%
\providecommand \@ifx [1]{%
 \ifx #1\expandafter \@firstoftwo
 \else \expandafter \@secondoftwo
 \fi
}%
\providecommand \natexlab [1]{#1}%
\providecommand \enquote  [1]{``#1''}%
\providecommand \bibnamefont  [1]{#1}%
\providecommand \bibfnamefont [1]{#1}%
\providecommand \citenamefont [1]{#1}%
\providecommand \href@noop [0]{\@secondoftwo}%
\providecommand \href [0]{\begingroup \@sanitize@url \@href}%
\providecommand \@href[1]{\@@startlink{#1}\@@href}%
\providecommand \@@href[1]{\endgroup#1\@@endlink}%
\providecommand \@sanitize@url [0]{\catcode `\\12\catcode `\$12\catcode
  `\&12\catcode `\#12\catcode `\^12\catcode `\_12\catcode `\%12\relax}%
\providecommand \@@startlink[1]{}%
\providecommand \@@endlink[0]{}%
\providecommand \url  [0]{\begingroup\@sanitize@url \@url }%
\providecommand \@url [1]{\endgroup\@href {#1}{\urlprefix }}%
\providecommand \urlprefix  [0]{URL }%
\providecommand \Eprint [0]{\href }%
\providecommand \doibase [0]{http://dx.doi.org/}%
\providecommand \selectlanguage [0]{\@gobble}%
\providecommand \bibinfo  [0]{\@secondoftwo}%
\providecommand \bibfield  [0]{\@secondoftwo}%
\providecommand \translation [1]{[#1]}%
\providecommand \BibitemOpen [0]{}%
\providecommand \bibitemStop [0]{}%
\providecommand \bibitemNoStop [0]{.\EOS\space}%
\providecommand \EOS [0]{\spacefactor3000\relax}%
\providecommand \BibitemShut  [1]{\csname bibitem#1\endcsname}%
\let\auto@bib@innerbib\@empty
\bibitem [{\citenamefont {Dubrulle}\ and\ \citenamefont
  {Frisch}(1991)}]{dubrulle1991}%
  \BibitemOpen
  \bibfield  {author} {\bibinfo {author} {\bibfnamefont {B.}~\bibnamefont
  {Dubrulle}}\ and\ \bibinfo {author} {\bibfnamefont {U.}~\bibnamefont
  {Frisch}},\ }\href {\doibase 10.1103/PhysRevA.43.5355} {\bibfield  {journal}
  {\bibinfo  {journal} {Phys. Rev. A}\ }\textbf {\bibinfo {volume} {43}},\
  \bibinfo {pages} {5355} (\bibinfo {year} {1991})}\BibitemShut {NoStop}%
\bibitem [{\citenamefont {Germano}\ \emph {et~al.}(1991)\citenamefont
  {Germano}, \citenamefont {Piomelli}, \citenamefont {Moin},\ and\
  \citenamefont {Cabot}}]{germano1991}%
  \BibitemOpen
  \bibfield  {author} {\bibinfo {author} {\bibfnamefont {M.}~\bibnamefont
  {Germano}}, \bibinfo {author} {\bibfnamefont {U.}~\bibnamefont {Piomelli}},
  \bibinfo {author} {\bibfnamefont {P.}~\bibnamefont {Moin}}, \ and\ \bibinfo
  {author} {\bibfnamefont {W.~H.}\ \bibnamefont {Cabot}},\ }\href {\doibase
  10.1063/1.857955} {\bibfield  {journal} {\bibinfo  {journal} {Physics of
  Fluids A: Fluid Dynamics}\ }\textbf {\bibinfo {volume} {3}},\ \bibinfo
  {pages} {1760} (\bibinfo {year} {1991})},\ \Eprint
  {http://arxiv.org/abs/https://doi.org/10.1063/1.857955}
  {https://doi.org/10.1063/1.857955} \BibitemShut {NoStop}%
\bibitem [{\citenamefont {Gama}\ \emph {et~al.}(1994)\citenamefont {Gama},
  \citenamefont {Vergassola},\ and\ \citenamefont {Frisch}}]{gama1994}%
  \BibitemOpen
  \bibfield  {author} {\bibinfo {author} {\bibfnamefont {S.}~\bibnamefont
  {Gama}}, \bibinfo {author} {\bibfnamefont {M.}~\bibnamefont {Vergassola}}, \
  and\ \bibinfo {author} {\bibfnamefont {U.}~\bibnamefont {Frisch}},\ }\href
  {\doibase 10.1017/S0022112094003459} {\bibfield  {journal} {\bibinfo
  {journal} {Journal of Fluid Mechanics}\ }\textbf {\bibinfo {volume} {260}},\
  \bibinfo {pages} {95–126} (\bibinfo {year} {1994})}\BibitemShut {NoStop}%
\bibitem [{\citenamefont {Frisch}(1987)}]{frisch1987lectures}%
  \BibitemOpen
  \bibfield  {author} {\bibinfo {author} {\bibfnamefont {U.}~\bibnamefont
  {Frisch}},\ }\href@noop {} {\bibfield  {journal} {\bibinfo  {journal}
  {Lecture Notes, NCAR-GTP Summer School June 1987}\ ,\ \bibinfo {pages} {219}}
  (\bibinfo {year} {1987})}\BibitemShut {NoStop}%
\bibitem [{\citenamefont {Frisch}(1995)}]{frisch1995turbulence}%
  \BibitemOpen
  \bibfield  {author} {\bibinfo {author} {\bibfnamefont {U.}~\bibnamefont
  {Frisch}},\ }\href@noop {} {\emph {\bibinfo {title} {Turbulence: the legacy
  of {A. N.} {Kolmogorov}}}}\ (\bibinfo  {publisher} {Cambridge University
  Press},\ \bibinfo {year} {1995})\BibitemShut {NoStop}%
\bibitem [{\citenamefont {Biferale}\ \emph
  {et~al.}(1995{\natexlab{a}})\citenamefont {Biferale}, \citenamefont
  {Crisanti}, \citenamefont {Vergassola},\ and\ \citenamefont
  {Vulpiani}}]{biferale1995}%
  \BibitemOpen
  \bibfield  {author} {\bibinfo {author} {\bibfnamefont {L.}~\bibnamefont
  {Biferale}}, \bibinfo {author} {\bibfnamefont {A.}~\bibnamefont {Crisanti}},
  \bibinfo {author} {\bibfnamefont {M.}~\bibnamefont {Vergassola}}, \ and\
  \bibinfo {author} {\bibfnamefont {A.}~\bibnamefont {Vulpiani}},\ }\href
  {\doibase 10.1063/1.868651} {\bibfield  {journal} {\bibinfo  {journal}
  {Physics of Fluids}\ }\textbf {\bibinfo {volume} {7}},\ \bibinfo {pages}
  {2725} (\bibinfo {year} {1995}{\natexlab{a}})},\ \Eprint
  {http://arxiv.org/abs/https://doi.org/10.1063/1.868651}
  {https://doi.org/10.1063/1.868651} \BibitemShut {NoStop}%
\bibitem [{\citenamefont {Mazzino}(1997)}]{mazzino1997}%
  \BibitemOpen
  \bibfield  {author} {\bibinfo {author} {\bibfnamefont {A.}~\bibnamefont
  {Mazzino}},\ }\href {\doibase 10.1103/PhysRevE.56.5500} {\bibfield  {journal}
  {\bibinfo  {journal} {Phys. Rev. E}\ }\textbf {\bibinfo {volume} {56}},\
  \bibinfo {pages} {5500} (\bibinfo {year} {1997})}\BibitemShut {NoStop}%
\bibitem [{\citenamefont {Castiglione}\ \emph {et~al.}(1998)\citenamefont
  {Castiglione}, \citenamefont {Crisanti}, \citenamefont {Mazzino},
  \citenamefont {Vergassola},\ and\ \citenamefont
  {Vulpiani}}]{castiglione1998}%
  \BibitemOpen
  \bibfield  {author} {\bibinfo {author} {\bibfnamefont {P.}~\bibnamefont
  {Castiglione}}, \bibinfo {author} {\bibfnamefont {A.}~\bibnamefont
  {Crisanti}}, \bibinfo {author} {\bibfnamefont {A.}~\bibnamefont {Mazzino}},
  \bibinfo {author} {\bibfnamefont {M.}~\bibnamefont {Vergassola}}, \ and\
  \bibinfo {author} {\bibfnamefont {A.}~\bibnamefont {Vulpiani}},\ }\href
  {\doibase 10.1088/0305-4470/31/35/002} {\bibfield  {journal} {\bibinfo
  {journal} {Journal of Physics A: Mathematical and General}\ }\textbf
  {\bibinfo {volume} {31}},\ \bibinfo {pages} {7197} (\bibinfo {year}
  {1998})}\BibitemShut {NoStop}%
\bibitem [{\citenamefont {Mazzino}\ \emph {et~al.}(2005)\citenamefont
  {Mazzino}, \citenamefont {Musacchio},\ and\ \citenamefont
  {Vulpiani}}]{mazzino2005}%
  \BibitemOpen
  \bibfield  {author} {\bibinfo {author} {\bibfnamefont {A.}~\bibnamefont
  {Mazzino}}, \bibinfo {author} {\bibfnamefont {S.}~\bibnamefont {Musacchio}},
  \ and\ \bibinfo {author} {\bibfnamefont {A.}~\bibnamefont {Vulpiani}},\
  }\href {\doibase 10.1103/PhysRevE.71.011113} {\bibfield  {journal} {\bibinfo
  {journal} {Phys. Rev. E}\ }\textbf {\bibinfo {volume} {71}},\ \bibinfo
  {pages} {011113} (\bibinfo {year} {2005})}\BibitemShut {NoStop}%
\bibitem [{\citenamefont {LeMone}\ \emph {et~al.}(2018)\citenamefont {LeMone},
  \citenamefont {Angevine}, \citenamefont {Bretherton}, \citenamefont {Chen},
  \citenamefont {Dudhia}, \citenamefont {Fedorovich}, \citenamefont {Katsaros},
  \citenamefont {Lenschow}, \citenamefont {Mahrt}, \citenamefont {Patton},
  \citenamefont {Sun}, \citenamefont {Tjernström},\ and\ \citenamefont
  {Weil}}]{lemone100years}%
  \BibitemOpen
  \bibfield  {author} {\bibinfo {author} {\bibfnamefont {M.~A.}\ \bibnamefont
  {LeMone}}, \bibinfo {author} {\bibfnamefont {W.~M.}\ \bibnamefont
  {Angevine}}, \bibinfo {author} {\bibfnamefont {C.~S.}\ \bibnamefont
  {Bretherton}}, \bibinfo {author} {\bibfnamefont {F.}~\bibnamefont {Chen}},
  \bibinfo {author} {\bibfnamefont {J.}~\bibnamefont {Dudhia}}, \bibinfo
  {author} {\bibfnamefont {E.}~\bibnamefont {Fedorovich}}, \bibinfo {author}
  {\bibfnamefont {K.~B.}\ \bibnamefont {Katsaros}}, \bibinfo {author}
  {\bibfnamefont {D.~H.}\ \bibnamefont {Lenschow}}, \bibinfo {author}
  {\bibfnamefont {L.}~\bibnamefont {Mahrt}}, \bibinfo {author} {\bibfnamefont
  {E.~G.}\ \bibnamefont {Patton}}, \bibinfo {author} {\bibfnamefont
  {J.}~\bibnamefont {Sun}}, \bibinfo {author} {\bibfnamefont {M.}~\bibnamefont
  {Tjernström}}, \ and\ \bibinfo {author} {\bibfnamefont {J.}~\bibnamefont
  {Weil}},\ }\href {\doibase 10.1175/AMSMONOGRAPHS-D-18-0013.1} {\bibfield
  {journal} {\bibinfo  {journal} {Meteorol. Monogr.}\ }\textbf {\bibinfo
  {volume} {59}},\ \bibinfo {pages} {9.1} (\bibinfo {year} {2018})},\ \Eprint
  {http://arxiv.org/abs/https://doi.org/10.1175/AMSMONOGRAPHS-D-18-0013.1}
  {https://doi.org/10.1175/AMSMONOGRAPHS-D-18-0013.1} \BibitemShut {NoStop}%
\bibitem [{\citenamefont {Raupach}\ and\ \citenamefont
  {Thom}(1981)}]{raupach1981}%
  \BibitemOpen
  \bibfield  {author} {\bibinfo {author} {\bibfnamefont {M.~R.}\ \bibnamefont
  {Raupach}}\ and\ \bibinfo {author} {\bibfnamefont {A.~S.}\ \bibnamefont
  {Thom}},\ }\href {\doibase 10.1146/annurev.fl.13.010181.000525} {\bibfield
  {journal} {\bibinfo  {journal} {Annu. Rev. Fluid Mech.}\ }\textbf {\bibinfo
  {volume} {13}},\ \bibinfo {pages} {97} (\bibinfo {year} {1981})},\ \Eprint
  {http://arxiv.org/abs/https://doi.org/10.1146/annurev.fl.13.010181.000525}
  {https://doi.org/10.1146/annurev.fl.13.010181.000525} \BibitemShut {NoStop}%
\bibitem [{\citenamefont {Poggi}\ \emph {et~al.}(2004)\citenamefont {Poggi},
  \citenamefont {Porporato}, \citenamefont {Ridolfi}, \citenamefont
  {Albertson},\ and\ \citenamefont {Katul}}]{poggi2004}%
  \BibitemOpen
  \bibfield  {author} {\bibinfo {author} {\bibfnamefont {D.}~\bibnamefont
  {Poggi}}, \bibinfo {author} {\bibfnamefont {A.}~\bibnamefont {Porporato}},
  \bibinfo {author} {\bibfnamefont {L.}~\bibnamefont {Ridolfi}}, \bibinfo
  {author} {\bibfnamefont {J.~D.}\ \bibnamefont {Albertson}}, \ and\ \bibinfo
  {author} {\bibfnamefont {G.~G.}\ \bibnamefont {Katul}},\ }\href {\doibase
  10.1023/B:BOUN.0000016576.05621.73} {\bibfield  {journal} {\bibinfo
  {journal} {Boundary-Layer Meteorol.}\ }\textbf {\bibinfo {volume} {111}},\
  \bibinfo {pages} {565} (\bibinfo {year} {2004})}\BibitemShut {NoStop}%
\bibitem [{\citenamefont {Di~Bernardino}\ \emph {et~al.}(2017)\citenamefont
  {Di~Bernardino}, \citenamefont {Monti}, \citenamefont {Leuzzi},\ and\
  \citenamefont {Querzoli}}]{DiBernardino2017}%
  \BibitemOpen
  \bibfield  {author} {\bibinfo {author} {\bibfnamefont {A.}~\bibnamefont
  {Di~Bernardino}}, \bibinfo {author} {\bibfnamefont {P.}~\bibnamefont
  {Monti}}, \bibinfo {author} {\bibfnamefont {G.}~\bibnamefont {Leuzzi}}, \
  and\ \bibinfo {author} {\bibfnamefont {G.}~\bibnamefont {Querzoli}},\ }\href
  {\doibase 10.1007/s10546-017-0278-6} {\bibfield  {journal} {\bibinfo
  {journal} {Boundary-Layer Meteorol.}\ }\textbf {\bibinfo {volume} {165}},\
  \bibinfo {pages} {251} (\bibinfo {year} {2017})}\BibitemShut {NoStop}%
\bibitem [{\citenamefont {Ghisalberti}\ and\ \citenamefont
  {Nepf}(2009)}]{ghisalberti_nepf_2009a}%
  \BibitemOpen
  \bibfield  {author} {\bibinfo {author} {\bibfnamefont {M.}~\bibnamefont
  {Ghisalberti}}\ and\ \bibinfo {author} {\bibfnamefont {H.}~\bibnamefont
  {Nepf}},\ }\href@noop {} {\bibfield  {journal} {\bibinfo  {journal}
  {{T}ransport in {P}orous {M}edia}\ }\textbf {\bibinfo {volume} {78}},\
  \bibinfo {pages} {309} (\bibinfo {year} {2009})}\BibitemShut {NoStop}%
\bibitem [{\citenamefont {Shnapp}\ \emph {et~al.}(2019)\citenamefont {Shnapp},
  \citenamefont {Shapira}, \citenamefont {Peri}, \citenamefont {Bohbot-Raviv},
  \citenamefont {Fattal},\ and\ \citenamefont {Liberzon}}]{shnapp2019extended}%
  \BibitemOpen
  \bibfield  {author} {\bibinfo {author} {\bibfnamefont {R.}~\bibnamefont
  {Shnapp}}, \bibinfo {author} {\bibfnamefont {E.}~\bibnamefont {Shapira}},
  \bibinfo {author} {\bibfnamefont {D.}~\bibnamefont {Peri}}, \bibinfo {author}
  {\bibfnamefont {Y.}~\bibnamefont {Bohbot-Raviv}}, \bibinfo {author}
  {\bibfnamefont {E.}~\bibnamefont {Fattal}}, \ and\ \bibinfo {author}
  {\bibfnamefont {A.}~\bibnamefont {Liberzon}},\ }\href@noop {} {\bibfield
  {journal} {\bibinfo  {journal} {Sci. Rep.}\ }\textbf {\bibinfo {volume}
  {9}},\ \bibinfo {pages} {7405} (\bibinfo {year} {2019})}\BibitemShut
  {NoStop}%
\bibitem [{\citenamefont {Zampogna}\ and\ \citenamefont
  {Bottaro}(2016)}]{zampogna_bottaro_2016}%
  \BibitemOpen
  \bibfield  {author} {\bibinfo {author} {\bibfnamefont {G.~A.}\ \bibnamefont
  {Zampogna}}\ and\ \bibinfo {author} {\bibfnamefont {A.}~\bibnamefont
  {Bottaro}},\ }\href {\doibase 10.1017/jfm.2016.66} {\bibfield  {journal}
  {\bibinfo  {journal} {Journal of Fluid Mechanics}\ }\textbf {\bibinfo
  {volume} {792}},\ \bibinfo {pages} {5–35} (\bibinfo {year}
  {2016})}\BibitemShut {NoStop}%
\bibitem [{\citenamefont {Zampogna}\ \emph {et~al.}(2016)\citenamefont
  {Zampogna}, \citenamefont {Pluvinage}, \citenamefont {Kourta},\ and\
  \citenamefont {Bottaro}}]{zampogna2016}%
  \BibitemOpen
  \bibfield  {author} {\bibinfo {author} {\bibfnamefont {G.~A.}\ \bibnamefont
  {Zampogna}}, \bibinfo {author} {\bibfnamefont {F.}~\bibnamefont {Pluvinage}},
  \bibinfo {author} {\bibfnamefont {A.}~\bibnamefont {Kourta}}, \ and\ \bibinfo
  {author} {\bibfnamefont {A.}~\bibnamefont {Bottaro}},\ }\href {\doibase
  10.1002/2016WR018915} {\bibfield  {journal} {\bibinfo  {journal} {Water
  Resources Research}\ }\textbf {\bibinfo {volume} {52}},\ \bibinfo {pages}
  {5421} (\bibinfo {year} {2016})},\ \Eprint
  {http://arxiv.org/abs/https://agupubs.onlinelibrary.wiley.com/doi/pdf/10.1002/2016WR018915}
  {https://agupubs.onlinelibrary.wiley.com/doi/pdf/10.1002/2016WR018915}
  \BibitemShut {NoStop}%
\bibitem [{\citenamefont {Monti}\ \emph {et~al.}(2019)\citenamefont {Monti},
  \citenamefont {Omidyeganeh},\ and\ \citenamefont {Pinelli}}]{monti2019PoF}%
  \BibitemOpen
  \bibfield  {author} {\bibinfo {author} {\bibfnamefont {A.}~\bibnamefont
  {Monti}}, \bibinfo {author} {\bibfnamefont {M.}~\bibnamefont {Omidyeganeh}},
  \ and\ \bibinfo {author} {\bibfnamefont {A.}~\bibnamefont {Pinelli}},\ }\href
  {\doibase 10.1063/1.5095770} {\bibfield  {journal} {\bibinfo  {journal}
  {Physics of Fluids}\ }\textbf {\bibinfo {volume} {31}},\ \bibinfo {pages}
  {065108} (\bibinfo {year} {2019})},\ \Eprint
  {http://arxiv.org/abs/https://doi.org/10.1063/1.5095770}
  {https://doi.org/10.1063/1.5095770} \BibitemShut {NoStop}%
\bibitem [{\citenamefont {Finnigan}(2000)}]{finnigan2000review}%
  \BibitemOpen
  \bibfield  {author} {\bibinfo {author} {\bibfnamefont {J.}~\bibnamefont
  {Finnigan}},\ }\href {\doibase 10.1146/annurev.fluid.32.1.519} {\bibfield
  {journal} {\bibinfo  {journal} {Annu. Rev. Fluid Mech.}\ }\textbf {\bibinfo
  {volume} {32}},\ \bibinfo {pages} {519} (\bibinfo {year} {2000})},\ \Eprint
  {http://arxiv.org/abs/https://doi.org/10.1146/annurev.fluid.32.1.519}
  {https://doi.org/10.1146/annurev.fluid.32.1.519} \BibitemShut {NoStop}%
\bibitem [{\citenamefont {Nepf}(2012)}]{nepf2012review}%
  \BibitemOpen
  \bibfield  {author} {\bibinfo {author} {\bibfnamefont {H.~M.}\ \bibnamefont
  {Nepf}},\ }\href {\doibase 10.1146/annurev-fluid-120710-101048} {\bibfield
  {journal} {\bibinfo  {journal} {Annual Review of Fluid Mechanics}\ }\textbf
  {\bibinfo {volume} {44}},\ \bibinfo {pages} {123} (\bibinfo {year} {2012})},\
  \Eprint
  {http://arxiv.org/abs/https://doi.org/10.1146/annurev-fluid-120710-101048}
  {https://doi.org/10.1146/annurev-fluid-120710-101048} \BibitemShut {NoStop}%
\bibitem [{\citenamefont {Galloway}\ and\ \citenamefont
  {Frisch}(1987)}]{galloway1987stability}%
  \BibitemOpen
  \bibfield  {author} {\bibinfo {author} {\bibfnamefont {D.}~\bibnamefont
  {Galloway}}\ and\ \bibinfo {author} {\bibfnamefont {U.}~\bibnamefont
  {Frisch}},\ }\href {\doibase 10.1017/S0022112087001952} {\bibfield  {journal}
  {\bibinfo  {journal} {Journal of Fluid Mechanics}\ }\textbf {\bibinfo
  {volume} {180}},\ \bibinfo {pages} {557–564} (\bibinfo {year}
  {1987})}\BibitemShut {NoStop}%
\bibitem [{\citenamefont {Podvigina}\ and\ \citenamefont
  {Pouquet}(1994)}]{podvigina1994}%
  \BibitemOpen
  \bibfield  {author} {\bibinfo {author} {\bibfnamefont {O.}~\bibnamefont
  {Podvigina}}\ and\ \bibinfo {author} {\bibfnamefont {A.}~\bibnamefont
  {Pouquet}},\ }\href {\doibase https://doi.org/10.1016/0167-2789(94)00031-X}
  {\bibfield  {journal} {\bibinfo  {journal} {Physica D: Nonlinear Phenomena}\
  }\textbf {\bibinfo {volume} {75}},\ \bibinfo {pages} {471 } (\bibinfo {year}
  {1994})}\BibitemShut {NoStop}%
\bibitem [{\citenamefont {du~Roure}\ \emph {et~al.}(2019)\citenamefont
  {du~Roure}, \citenamefont {Lindner}, \citenamefont {Nazockdast},\ and\
  \citenamefont {Shelley}}]{duroure2019review}%
  \BibitemOpen
  \bibfield  {author} {\bibinfo {author} {\bibfnamefont {O.}~\bibnamefont
  {du~Roure}}, \bibinfo {author} {\bibfnamefont {A.}~\bibnamefont {Lindner}},
  \bibinfo {author} {\bibfnamefont {E.~N.}\ \bibnamefont {Nazockdast}}, \ and\
  \bibinfo {author} {\bibfnamefont {M.~J.}\ \bibnamefont {Shelley}},\ }\href
  {\doibase 10.1146/annurev-fluid-122316-045153} {\bibfield  {journal}
  {\bibinfo  {journal} {Annu. Rev. Fluid Mech.}\ }\textbf {\bibinfo {volume}
  {51}},\ \bibinfo {pages} {539} (\bibinfo {year} {2019})}\BibitemShut
  {NoStop}%
\bibitem [{\citenamefont {Rosti}\ \emph {et~al.}(2018)\citenamefont {Rosti},
  \citenamefont {Banaei}, \citenamefont {Brandt},\ and\ \citenamefont
  {Mazzino}}]{rosti2018flexible}%
  \BibitemOpen
  \bibfield  {author} {\bibinfo {author} {\bibfnamefont {M.~E.}\ \bibnamefont
  {Rosti}}, \bibinfo {author} {\bibfnamefont {A.~A.}\ \bibnamefont {Banaei}},
  \bibinfo {author} {\bibfnamefont {L.}~\bibnamefont {Brandt}}, \ and\ \bibinfo
  {author} {\bibfnamefont {A.}~\bibnamefont {Mazzino}},\ }\href {\doibase
  10.1103/PhysRevLett.121.044501} {\bibfield  {journal} {\bibinfo  {journal}
  {Phys. Rev. Lett.}\ }\textbf {\bibinfo {volume} {121}},\ \bibinfo {pages}
  {044501} (\bibinfo {year} {2018})}\BibitemShut {NoStop}%
\bibitem [{\citenamefont {Rosti}\ \emph {et~al.}(2019)\citenamefont {Rosti},
  \citenamefont {Olivieri}, \citenamefont {Banaei}, \citenamefont {Brandt},\
  and\ \citenamefont {Mazzino}}]{rosti2019flowing}%
  \BibitemOpen
  \bibfield  {author} {\bibinfo {author} {\bibfnamefont {M.~E.}\ \bibnamefont
  {Rosti}}, \bibinfo {author} {\bibfnamefont {S.}~\bibnamefont {Olivieri}},
  \bibinfo {author} {\bibfnamefont {A.~A.}\ \bibnamefont {Banaei}}, \bibinfo
  {author} {\bibfnamefont {L.}~\bibnamefont {Brandt}}, \ and\ \bibinfo {author}
  {\bibfnamefont {A.}~\bibnamefont {Mazzino}},\ }\href {\doibase
  10.1007/s11012-019-00997-2} {\bibfield  {journal} {\bibinfo  {journal}
  {Meccanica}\ } (\bibinfo {year} {2019}),\
  10.1007/s11012-019-00997-2}\BibitemShut {NoStop}%
\bibitem [{\citenamefont {Banaei}\ \emph {et~al.}(2020)\citenamefont {Banaei},
  \citenamefont {Rosti},\ and\ \citenamefont {Brandt}}]{banaei2019numerical}%
  \BibitemOpen
  \bibfield  {author} {\bibinfo {author} {\bibfnamefont {A.~A.}\ \bibnamefont
  {Banaei}}, \bibinfo {author} {\bibfnamefont {M.~E.}\ \bibnamefont {Rosti}}, \
  and\ \bibinfo {author} {\bibfnamefont {L.}~\bibnamefont {Brandt}},\ }\href
  {\doibase 10.1017/jfm.2019.794} {\bibfield  {journal} {\bibinfo  {journal}
  {Journal of Fluid Mechanics}\ }\textbf {\bibinfo {volume} {882}},\ \bibinfo
  {pages} {A5} (\bibinfo {year} {2020})}\BibitemShut {NoStop}%
\bibitem [{\citenamefont {Cavaiola}\ \emph {et~al.}(2019)\citenamefont
  {Cavaiola}, \citenamefont {Olivieri},\ and\ \citenamefont
  {Mazzino}}]{cavaiola2019assembly}%
  \BibitemOpen
  \bibfield  {author} {\bibinfo {author} {\bibfnamefont {M.}~\bibnamefont
  {Cavaiola}}, \bibinfo {author} {\bibfnamefont {S.}~\bibnamefont {Olivieri}},
  \ and\ \bibinfo {author} {\bibfnamefont {A.}~\bibnamefont {Mazzino}},\
  }\href@noop {} {\bibfield  {journal} {\bibinfo  {journal} {arXiv preprint
  arXiv:1908.04072}\ } (\bibinfo {year} {2019})}\BibitemShut {NoStop}%
\bibitem [{\citenamefont {Huang}\ \emph {et~al.}(2007)\citenamefont {Huang},
  \citenamefont {Shin},\ and\ \citenamefont {Sung}}]{huang_shin_sung_2007a}%
  \BibitemOpen
  \bibfield  {author} {\bibinfo {author} {\bibfnamefont {W.-X.}\ \bibnamefont
  {Huang}}, \bibinfo {author} {\bibfnamefont {S.~J.}\ \bibnamefont {Shin}}, \
  and\ \bibinfo {author} {\bibfnamefont {H.~J.}\ \bibnamefont {Sung}},\ }\href
  {\doibase https://doi.org/10.1016/j.jcp.2007.07.002} {\bibfield  {journal}
  {\bibinfo  {journal} {J. Comput. Phys.}\ }\textbf {\bibinfo {volume} {226}},\
  \bibinfo {pages} {2206 } (\bibinfo {year} {2007})}\BibitemShut {NoStop}%
\bibitem [{\citenamefont {Roma}\ \emph {et~al.}(1999)\citenamefont {Roma},
  \citenamefont {Peskin},\ and\ \citenamefont {Berger}}]{roma1999}%
  \BibitemOpen
  \bibfield  {author} {\bibinfo {author} {\bibfnamefont {A.~M.}\ \bibnamefont
  {Roma}}, \bibinfo {author} {\bibfnamefont {C.~S.}\ \bibnamefont {Peskin}}, \
  and\ \bibinfo {author} {\bibfnamefont {M.~J.}\ \bibnamefont {Berger}},\
  }\href {\doibase https://doi.org/10.1006/jcph.1999.6293} {\bibfield
  {journal} {\bibinfo  {journal} {J. Comput. Phys.}\ }\textbf {\bibinfo
  {volume} {153}},\ \bibinfo {pages} {509 } (\bibinfo {year}
  {1999})}\BibitemShut {NoStop}%
\bibitem [{\citenamefont {Rosti}\ and\ \citenamefont
  {Brandt}(2017)}]{rosti_brandt_2017a}%
  \BibitemOpen
  \bibfield  {author} {\bibinfo {author} {\bibfnamefont {M.~E.}\ \bibnamefont
  {Rosti}}\ and\ \bibinfo {author} {\bibfnamefont {L.}~\bibnamefont {Brandt}},\
  }\href@noop {} {\bibfield  {journal} {\bibinfo  {journal} {{J}ournal of
  {F}luid {M}echanics}\ }\textbf {\bibinfo {volume} {830}},\ \bibinfo {pages}
  {708} (\bibinfo {year} {2017})}\BibitemShut {NoStop}%
\bibitem [{\citenamefont {Dombre}\ \emph {et~al.}(1986)\citenamefont {Dombre},
  \citenamefont {Frisch}, \citenamefont {Greene}, \citenamefont {H\'enon},
  \citenamefont {Mehr},\ and\ \citenamefont {Soward}}]{dombre1986chaotic}%
  \BibitemOpen
  \bibfield  {author} {\bibinfo {author} {\bibfnamefont {T.}~\bibnamefont
  {Dombre}}, \bibinfo {author} {\bibfnamefont {U.}~\bibnamefont {Frisch}},
  \bibinfo {author} {\bibfnamefont {J.~M.}\ \bibnamefont {Greene}}, \bibinfo
  {author} {\bibfnamefont {M.}~\bibnamefont {H\'enon}}, \bibinfo {author}
  {\bibfnamefont {A.}~\bibnamefont {Mehr}}, \ and\ \bibinfo {author}
  {\bibfnamefont {A.~M.}\ \bibnamefont {Soward}},\ }\href {\doibase
  10.1017/S0022112086002859} {\bibfield  {journal} {\bibinfo  {journal} {J.
  Fluid Mech.}\ }\textbf {\bibinfo {volume} {167}},\ \bibinfo {pages}
  {353–391} (\bibinfo {year} {1986})}\BibitemShut {NoStop}%
\bibitem [{\citenamefont {Arnold}(1965)}]{arnold1965topologie}%
  \BibitemOpen
  \bibfield  {author} {\bibinfo {author} {\bibfnamefont {V.~I.}\ \bibnamefont
  {Arnold}},\ }in\ \href@noop {} {\emph {\bibinfo {booktitle} {Vladimir I.
  Arnold-Collected Works}}}\ (\bibinfo  {publisher} {Springer},\ \bibinfo
  {year} {1965})\ pp.\ \bibinfo {pages} {15--18}\BibitemShut {NoStop}%
\bibitem [{\citenamefont {Zhao}\ \emph {et~al.}(1993)\citenamefont {Zhao},
  \citenamefont {Kwek}, \citenamefont {Li},\ and\ \citenamefont
  {Huang}}]{zhao1993chaotic}%
  \BibitemOpen
  \bibfield  {author} {\bibinfo {author} {\bibfnamefont {X.}~\bibnamefont
  {Zhao}}, \bibinfo {author} {\bibfnamefont {K.}~\bibnamefont {Kwek}}, \bibinfo
  {author} {\bibfnamefont {J.}~\bibnamefont {Li}}, \ and\ \bibinfo {author}
  {\bibfnamefont {K.}~\bibnamefont {Huang}},\ }\href {\doibase 10.1137/0153005}
  {\bibfield  {journal} {\bibinfo  {journal} {SIAM Journal on Applied
  Mathematics}\ }\textbf {\bibinfo {volume} {53}},\ \bibinfo {pages} {71}
  (\bibinfo {year} {1993})},\ \Eprint
  {http://arxiv.org/abs/https://doi.org/10.1137/0153005}
  {https://doi.org/10.1137/0153005} \BibitemShut {NoStop}%
\bibitem [{\citenamefont {Biferale}\ \emph
  {et~al.}(1995{\natexlab{b}})\citenamefont {Biferale}, \citenamefont
  {Crisanti}, \citenamefont {Vergassola},\ and\ \citenamefont
  {Vulpiani}}]{biferale1995eddy}%
  \BibitemOpen
  \bibfield  {author} {\bibinfo {author} {\bibfnamefont {L.}~\bibnamefont
  {Biferale}}, \bibinfo {author} {\bibfnamefont {A.}~\bibnamefont {Crisanti}},
  \bibinfo {author} {\bibfnamefont {M.}~\bibnamefont {Vergassola}}, \ and\
  \bibinfo {author} {\bibfnamefont {A.}~\bibnamefont {Vulpiani}},\ }\href
  {\doibase 10.1063/1.868651} {\bibfield  {journal} {\bibinfo  {journal} {Phys.
  Fluids}\ }\textbf {\bibinfo {volume} {7}},\ \bibinfo {pages} {2725} (\bibinfo
  {year} {1995}{\natexlab{b}})}\BibitemShut {NoStop}%
\bibitem [{\citenamefont {Bottaro}(2019)}]{bottaro2019review}%
  \BibitemOpen
  \bibfield  {author} {\bibinfo {author} {\bibfnamefont {A.}~\bibnamefont
  {Bottaro}},\ }\href {\doibase 10.1017/jfm.2019.607} {\bibfield  {journal}
  {\bibinfo  {journal} {Journal of Fluid Mechanics}\ }\textbf {\bibinfo
  {volume} {877}},\ \bibinfo {pages} {P1} (\bibinfo {year} {2019})}\BibitemShut
  {NoStop}%
\bibitem [{\citenamefont {Meshalkin}\ and\ \citenamefont
  {Sinai}(1961)}]{meshalkin1961}%
  \BibitemOpen
  \bibfield  {author} {\bibinfo {author} {\bibfnamefont {L.}~\bibnamefont
  {Meshalkin}}\ and\ \bibinfo {author} {\bibfnamefont {I.}~\bibnamefont
  {Sinai}},\ }\href {\doibase https://doi.org/10.1016/0021-8928(62)90149-1}
  {\bibfield  {journal} {\bibinfo  {journal} {Journal of Applied Mathematics
  and Mechanics}\ }\textbf {\bibinfo {volume} {25}},\ \bibinfo {pages} {1700 }
  (\bibinfo {year} {1961})}\BibitemShut {NoStop}%
\bibitem [{\citenamefont {Thess}(1992)}]{thess1992}%
  \BibitemOpen
  \bibfield  {author} {\bibinfo {author} {\bibfnamefont {A.}~\bibnamefont
  {Thess}},\ }\href {\doibase 10.1063/1.858415} {\bibfield  {journal} {\bibinfo
   {journal} {Physics of Fluids A: Fluid Dynamics}\ }\textbf {\bibinfo {volume}
  {4}},\ \bibinfo {pages} {1385} (\bibinfo {year} {1992})},\ \Eprint
  {http://arxiv.org/abs/https://doi.org/10.1063/1.858415}
  {https://doi.org/10.1063/1.858415} \BibitemShut {NoStop}%
\bibitem [{\citenamefont {Boffetta}\ \emph
  {et~al.}(2005{\natexlab{a}})\citenamefont {Boffetta}, \citenamefont {Celani},
  \citenamefont {Mazzino}, \citenamefont {Puliafito},\ and\ \citenamefont
  {Vergassola}}]{boffetta2005viscoelastic}%
  \BibitemOpen
  \bibfield  {author} {\bibinfo {author} {\bibfnamefont {G.}~\bibnamefont
  {Boffetta}}, \bibinfo {author} {\bibfnamefont {A.}~\bibnamefont {Celani}},
  \bibinfo {author} {\bibfnamefont {A.}~\bibnamefont {Mazzino}}, \bibinfo
  {author} {\bibfnamefont {A.}~\bibnamefont {Puliafito}}, \ and\ \bibinfo
  {author} {\bibfnamefont {M.}~\bibnamefont {Vergassola}},\ }\href {\doibase
  10.1017/S0022112004002423} {\bibfield  {journal} {\bibinfo  {journal}
  {Journal of Fluid Mechanics}\ }\textbf {\bibinfo {volume} {523}},\ \bibinfo
  {pages} {161–170} (\bibinfo {year} {2005}{\natexlab{a}})}\BibitemShut
  {NoStop}%
\bibitem [{\citenamefont {Boffetta}\ \emph
  {et~al.}(2005{\natexlab{b}})\citenamefont {Boffetta}, \citenamefont
  {Celani},\ and\ \citenamefont {Mazzino}}]{boffetta2005drag}%
  \BibitemOpen
  \bibfield  {author} {\bibinfo {author} {\bibfnamefont {G.}~\bibnamefont
  {Boffetta}}, \bibinfo {author} {\bibfnamefont {A.}~\bibnamefont {Celani}}, \
  and\ \bibinfo {author} {\bibfnamefont {A.}~\bibnamefont {Mazzino}},\ }\href
  {\doibase 10.1103/PhysRevE.71.036307} {\bibfield  {journal} {\bibinfo
  {journal} {Phys. Rev. E}\ }\textbf {\bibinfo {volume} {71}},\ \bibinfo
  {pages} {036307} (\bibinfo {year} {2005}{\natexlab{b}})}\BibitemShut
  {NoStop}%
\bibitem [{\citenamefont {Bistagnino}\ \emph {et~al.}(2007)\citenamefont
  {Bistagnino}, \citenamefont {Boffetta}, \citenamefont {Celani}, \citenamefont
  {Mazzino}, \citenamefont {Puliafito},\ and\ \citenamefont
  {Vergassola}}]{bistagnino2007nonlinear}%
  \BibitemOpen
  \bibfield  {author} {\bibinfo {author} {\bibfnamefont {A.}~\bibnamefont
  {Bistagnino}}, \bibinfo {author} {\bibfnamefont {G.}~\bibnamefont
  {Boffetta}}, \bibinfo {author} {\bibfnamefont {A.}~\bibnamefont {Celani}},
  \bibinfo {author} {\bibfnamefont {A.}~\bibnamefont {Mazzino}}, \bibinfo
  {author} {\bibfnamefont {A.}~\bibnamefont {Puliafito}}, \ and\ \bibinfo
  {author} {\bibfnamefont {M.}~\bibnamefont {Vergassola}},\ }\href {\doibase
  10.1017/S0022112007007859} {\bibfield  {journal} {\bibinfo  {journal}
  {Journal of Fluid Mechanics}\ }\textbf {\bibinfo {volume} {590}},\ \bibinfo
  {pages} {61–80} (\bibinfo {year} {2007})}\BibitemShut {NoStop}%
\bibitem [{\citenamefont {Tithof}\ \emph {et~al.}(2017)\citenamefont {Tithof},
  \citenamefont {Suri}, \citenamefont {Pallantla}, \citenamefont {Grigoriev},\
  and\ \citenamefont {Schatz}}]{tithof2017bifurcations}%
  \BibitemOpen
  \bibfield  {author} {\bibinfo {author} {\bibfnamefont {J.}~\bibnamefont
  {Tithof}}, \bibinfo {author} {\bibfnamefont {B.}~\bibnamefont {Suri}},
  \bibinfo {author} {\bibfnamefont {R.~K.}\ \bibnamefont {Pallantla}}, \bibinfo
  {author} {\bibfnamefont {R.~O.}\ \bibnamefont {Grigoriev}}, \ and\ \bibinfo
  {author} {\bibfnamefont {M.~F.}\ \bibnamefont {Schatz}},\ }\href {\doibase
  10.1017/jfm.2017.553} {\bibfield  {journal} {\bibinfo  {journal} {Journal of
  Fluid Mechanics}\ }\textbf {\bibinfo {volume} {828}},\ \bibinfo {pages}
  {837–866} (\bibinfo {year} {2017})}\BibitemShut {NoStop}%
\bibitem [{\citenamefont {Eswaran}\ and\ \citenamefont
  {Pope}(1988)}]{eswaran1988forcing}%
  \BibitemOpen
  \bibfield  {author} {\bibinfo {author} {\bibfnamefont {V.}~\bibnamefont
  {Eswaran}}\ and\ \bibinfo {author} {\bibfnamefont {S.}~\bibnamefont {Pope}},\
  }\href {\doibase https://doi.org/10.1016/0045-7930(88)90013-8} {\bibfield
  {journal} {\bibinfo  {journal} {Computers \& Fluids}\ }\textbf {\bibinfo
  {volume} {16}},\ \bibinfo {pages} {257 } (\bibinfo {year}
  {1988})}\BibitemShut {NoStop}%
\bibitem [{\citenamefont {Pope}(2000)}]{pope2000turbulent}%
  \BibitemOpen
  \bibfield  {author} {\bibinfo {author} {\bibfnamefont {S.~B.}\ \bibnamefont
  {Pope}},\ }\href@noop {} {\emph {\bibinfo {title} {Turbulent Flows}}}\
  (\bibinfo  {publisher} {Cambridge University Press},\ \bibinfo {year}
  {2000})\BibitemShut {NoStop}%
\end{thebibliography}%

\end{document}